\documentclass[11pt, A4paper]{article}
\usepackage[utf8]{inputenc}
\usepackage[margin=1in]{geometry}
\usepackage[dvipsnames]{xcolor}
\usepackage{times}
\usepackage{amsmath}
\usepackage{amsfonts}
\usepackage{authblk}
\usepackage{graphicx, subfigure}
\usepackage{indentfirst}
\usepackage{amsthm}
\usepackage{lscape}
\usepackage[title,toc,titletoc]{appendix}

\usepackage{hyperref}
\hypersetup{colorlinks, allcolors=blue}
\usepackage{amssymb, bbm, bm}
\usepackage{setspace}
\usepackage{enumerate}
\usepackage[ruled,vlined]{algorithm2e}
\usepackage{booktabs, array}
\usepackage{siunitx} 
\usepackage{color, soul, colortbl}
\definecolor{LightCyan}{rgb}{0.88,1,1}

\SetKwRepeat{Do}{do}{while}
\theoremstyle{plain}

\numberwithin{theorem}{section}
\numberwithin{equation}{section}

\newcommand{\comment}[1]{}
\allowdisplaybreaks
\usepackage{array}
\newcolumntype{H}{>{\setbox0=\hbox\bgroup}c<{\egroup}@{}}
\usepackage{multirow}
\usepackage{circuitikz}
\usepackage{natbib}

\title{On the Impact of Spatial Covariance Matrix Ordering on Tile Low-Rank Estimation of  Mat\'ern Parameters}
\author{Sihan Chen$^{*}$, Sameh Abdulah$^{\dagger}$, Ying Sun$^{*, \dagger}$ and Marc G. Genton$^{*, \dagger}$ \\
\small{$^*$\textit{Statistics Program, King Abdullah University of Science and Technology}} \\
\small{$^\dagger$\textit{Extreme Computing Research Center (ECRC), King Abdullah University of Science and Technology}} \\
\small{\{sihan.chen, sameh.abdulah, ying.sun, marc.genton\}@kaust.edu.sa}}

\begin{document}

\maketitle

\begin{abstract}
Spatial statistical modeling and prediction involve generating and manipulating an $n \times n$ symmetric positive definite covariance matrix, where $n$ denotes the number of spatial locations. However, when $n$ is large, processing this covariance matrix using traditional methods becomes prohibitive. Thus, coupling parallel processing with approximation can be an elegant solution to this challenge by relying on parallel solvers that deal with the matrix as a set of small tiles instead of the full structure. Each processing unit can process a single tile, allowing better performance. The approximation can also be performed at the tile level for better compression and faster execution. The Tile Low-Rank (TLR) approximation, a tile-based approximation algorithm, has recently been used in spatial statistics applications. However, the quality of TLR algorithms mainly relies on ordering the matrix elements. This order can impact the compression quality and, therefore, the efficiency of the underlying linear solvers, which highly depends on the individual ranks of each tile. Thus, herein, we aim to investigate the accuracy and performance of some existing ordering algorithms that are used to order the geospatial locations before generating the spatial covariance matrix. Furthermore, we highlight the pros and cons of each ordering algorithm in the context of spatial statistics applications and give hints to practitioners on how to choose the ordering algorithm carefully. We assess the quality of the compression and the accuracy of the statistical parameter estimates of the Mat\'ern covariance function using TLR approximation under various ordering algorithms and settings of correlations.
\end{abstract}







\section{Introduction}

Spatial statistics is an important branch of statistics that has applications in various research fields, for instance, environmental science (\citealp{sun2015matern}), economics (\citealp{elhorst2021cross}), epidemiology (\citealp{moraga2011detection}), and neuroscience (\citealp{ombao2008spatio}), to name but a few. A common way to model spatial data is to consider them as realizations of a Gaussian random field. Suppose we have $n$ spatial locations $\mathbf{s}_1, \dots, \mathbf{s}_n\in\mathbb{R}^d$ for some $d\in\mathbb{Z}^+$, with $n\in\mathbb{Z}^+$. Denote the observations at these $n$ locations by $\mathbf{Z}=\{Z(\mathbf{s}_1), \dots, Z(\mathbf{s}_n)\}^\top$. We assume that the distribution of these observations is jointly Gaussian with mean $\mathbb{E}\{Z(\mathbf{s})\}=\mu(\mathbf{s})$ and covariance parametrized by $\boldsymbol{\theta}\in\mathbb{R}^q$ for some $q\in\mathbb{Z}^+$: $\text{Cov}\{Z(\mathbf{s}_1), Z(\mathbf{s}_2)\}=C(\mathbf{s}_1, \mathbf{s}_2;\boldsymbol{\theta})$. There are many valid parametric models established for the covariance function $C(\mathbf{s}_1, \mathbf{s}_2;\boldsymbol{\theta})$; see, e.g.,  \cite{gneiting2006geostatistical} and \cite{chen2021space}. 

Maximum Likelihood Estimation (MLE) is an essential technique for parameter estimation in geospatial data modeling. It hinges on maximizing a likelihood function, which measures how accurately the model reflects observed data. The MLE methodology entails constructing an $n \times n$ covariance matrix,  $\boldsymbol{\Sigma}(\boldsymbol{\theta})$, pivotal in defining the correlations between observations at different spatial locations across single or multiple time slots. Under the setting of a Gaussian random field with a single time slot, we form the covariance matrix $\boldsymbol{\Sigma}(\boldsymbol{\theta})$ by letting 
\begin{equation}
    \boldsymbol{\Sigma}_{i, j}(\boldsymbol{\theta})=C(\mathbf{s}_i, \mathbf{s}_j;\boldsymbol{\theta}),
    \label{covfunc}
\end{equation}
where $\boldsymbol{\Sigma}_{i, j}(\boldsymbol{\theta})$ denotes the $(i,j)$-th entry of $\boldsymbol{\Sigma}(\boldsymbol{\theta})$ for any $1\leq i, j\leq n$. Thus, we have the following expression of the Gaussian log-likelihood function for the observation vector $\mathbf{Z}$: 
\begin{equation}
    \ell(\boldsymbol{\theta})=-\frac{n}{2}\log(2\pi)-\frac{1}{2}\log|\mathbf{\Sigma}(\boldsymbol{\theta})|-\frac{1}{2}\mathbf{Z}^\top\mathbf{\Sigma}(\boldsymbol{\theta})^{-1}\mathbf{Z}.
    \label{loglik}
\end{equation}
Here $|\mathbf{\Sigma}(\boldsymbol{\theta})|$ denotes the determinant of $\mathbf{\Sigma}(\boldsymbol{\theta})$, and the MLE is obtained by maximizing the log-likelihood function $\ell(\boldsymbol{\theta})$ with respect to $\boldsymbol{\theta}$. 

However, as we can see from the expression, evaluating the function $\ell(\boldsymbol{\theta})$ requires computing the inverse of the covariance matrix $\mathbf{\Sigma}(\boldsymbol{\theta})$, with the time complexity $O(n^3)$. Although estimating the parameter $\boldsymbol{\theta}$ using the MLE is a classical way to understand the correlation structure of spatial data, the cubic computational complexity associated with the MLE renders its dense computation impractical with large $n$. This challenge is amplified by the availability of large-scale spatial data, where the location count can soar into millions or even hundreds of millions, particularly in cases involving high-resolution data. 

Consequently, recent research has focused on developing advanced approximation methods capable of handling large geospatial datasets while maintaining acceptable accuracy in the modeling process. These methods aim to offer more efficient alternatives for working with extensive spatial data without significantly compromising the quality of the results. For example, the covariance tapering method (\citealp{kaufman2008covariance}) sparsifies the covariance matrices by ignoring the correlations between locations with large distances and setting them to zero to accelerate the computation; \cite{cressie2008fixed} proposed a method that uses several non-stationary covariance functions to perform spatial prediction for large-scale spatial data; \cite{banerjee2008gaussian} proposed the Gaussian Predictive Processes (GPP), which projects the original problem to a subspace containing a set of spatial locations, to reduce the dimensionality of the spatial covariance matrix; the Mixed-Precision (MP) method (\citealp{cao2022reshaping}) aims to accelerate the computation by keeping the most important values with the highest level of precision, while truncating the rest of the values to lower levels of precision, so that the computation can be faster without affecting the accuracy too much. Some other approximation methods can also be found in the literature, such as Gaussian Markov random field approximations (\citealp{rue2002fitting} \& \citealp{rue2005gaussian}), Kalman filtering (\citealp{sinopoli2004kalman}) and low-rank splines (\citealp{kim2004smoothing}), to name but a few. \cite{sun2012geostatistics} provide a comprehensive overview of approximation methods on large-scale spatial data. 

Tile Low-Rank (TLR) approximation (\citealp{akbudak2017tile}) is one of the novel approximation methods. It employs low-rank approximation to the covariance matrix and facilitates parallel processing via a task-based parallelism mechanism. This approach significantly accelerates the evaluation of the likelihood function for a large number of locations on modern parallel hardware architectures where these task-based algorithms are optimized. In the TLR framework, low-rank compression is applied on individual tiles rather than the entire matrix. This strategy enables the distribution of both compression and computational tasks across various processing units, e.g., CPUs or GPUs, enhancing efficiency and scalability. The TLR approximation approach exploits the data sparsity of a given covariance matrix by compressing the off-diagonal tiles up to a user-defined accuracy threshold. In TLR approximation, the diagonal and off-diagonal tiles of the covariance matrices are stored and processed differently. The maximum rank among all the off-diagonal tiles is decisive on the performance, where smaller ranks lead to faster computations and less memory consumption. Since the approximation is applied at the tile level, ordering the covariance matrix elements can impact the compression level per tile. Therefore, one of the main requirements of TLR approximation is to spatially order the locations so that more correlated locations are stored together to allow better compression at the tile level.

Our work explores the accuracy and performance of geospatial data modeling using TLR approximation under varying orderings of locations. We use the {\em ExaGeoStat} software as our tool, as introduced in \cite{exageostat}. The ranks of the off-diagonal tiles can be affected by how we order the locations of the spatial data. In the literature, several multi-dimensional ordering methods have been proposed to sort the elements in a given $n\times n$-dimensional covariance matrix. Herein, we implemented several multi-dimensional ordering algorithms and assessed their quality for different covariance functions through an empirical study. We assess and compare the performance of TLR approximation method in estimating the Mat\'ern covariance parameters using orderings with the Morton curve (\citealp{morton}), the Hilbert curve (\citealp{hilbert}), the $k$-dimensional ($k$-d) tree (\citealp{kdtree}), as well as the Maximum-Minimum Degree (MMD) Algorithm (\citealp{guinness}), the Reversed Cuthill-McKee (RCM) Algorithm (\citealp{rcm}), and the Minimum Degree Algorithm (\citealp{george}), which we will describe in detail. 

The rest of this paper is organized as follows. Section~\ref{section:tlr} provides an overview of the TLR approximation method. Section~\ref{section:ordering} details the various ordering algorithms implemented in \textit{ExaGeoStat}, along with the experiments conducted using them. Section~\ref{section:model} introduces the statistical models under consideration, which describe the parametrization of the covariance matrix and the log-likelihood function. Section~\ref{section:exp} presents our experimental results, analyzing the performance of parameter estimation using the TLR algorithm and various ordering algorithms. This analysis focuses on estimation accuracy, ranks of the off-diagonal tiles in the covariance matrices, and execution time for Cholesky factorization on these matrices. An application to soil moisture data is also demonstrated, comparing results obtained with different ordering algorithms. The conclusion and discussion are provided in Section~\ref{section:con}.

\section{Tile Low-Rank (TLR) Approximation}
\label{section:tlr}

Parallel processing, an advanced computing paradigm, leverages modern parallel architectures to accelerate computation. This is achieved by employing multiple processing units simultaneously executing different parts of a single algorithm. In most of the existing parallel linear algebra libraries, task-based parallelism is a prevalent approach for enhancing the efficiency of linear solvers. This method involves dividing the computational workload into discrete tasks that can be executed simultaneously. The central strategy is to divide the target matrix into smaller tiles, allowing each processing unit to handle a specific tile independently. The algorithm is thus conceptualized as a Directed Acyclic Graph (DAG), where each node represents an individual task, and the connecting arrows indicate task dependencies. By utilizing runtime system libraries like OmpSs (\citealp{duran2011ompss}), StarPU (\citealp{augonnet2009starpu}), Charm++ (\citealp{kale1993charm++}), PaRSEC (\citealp{bosilca2013parsec}), and Kokkos (\citealp{edwards2014kokkos}), a scheduler adeptly assigns work to various processing units, ensuring no dependencies are violated during execution.

In \cite{akbudak2017tile}, the proposal of the tile low-rank (TLR) approach for compressing the covariance matrix specifically in climate/weather applications was presented. This study aimed to synergize fast, parallel processing, task-based linear algebra solvers with approximation techniques on manycore systems. Its application and evaluation in spatial statistics have been discussed in \cite{abdulah2018parallel} with accuracy evaluation for synthetic and real datasets. 

\begin{figure}[t!]
    \centering
    \vspace{-1.6cm}
    \includegraphics[width=\linewidth]{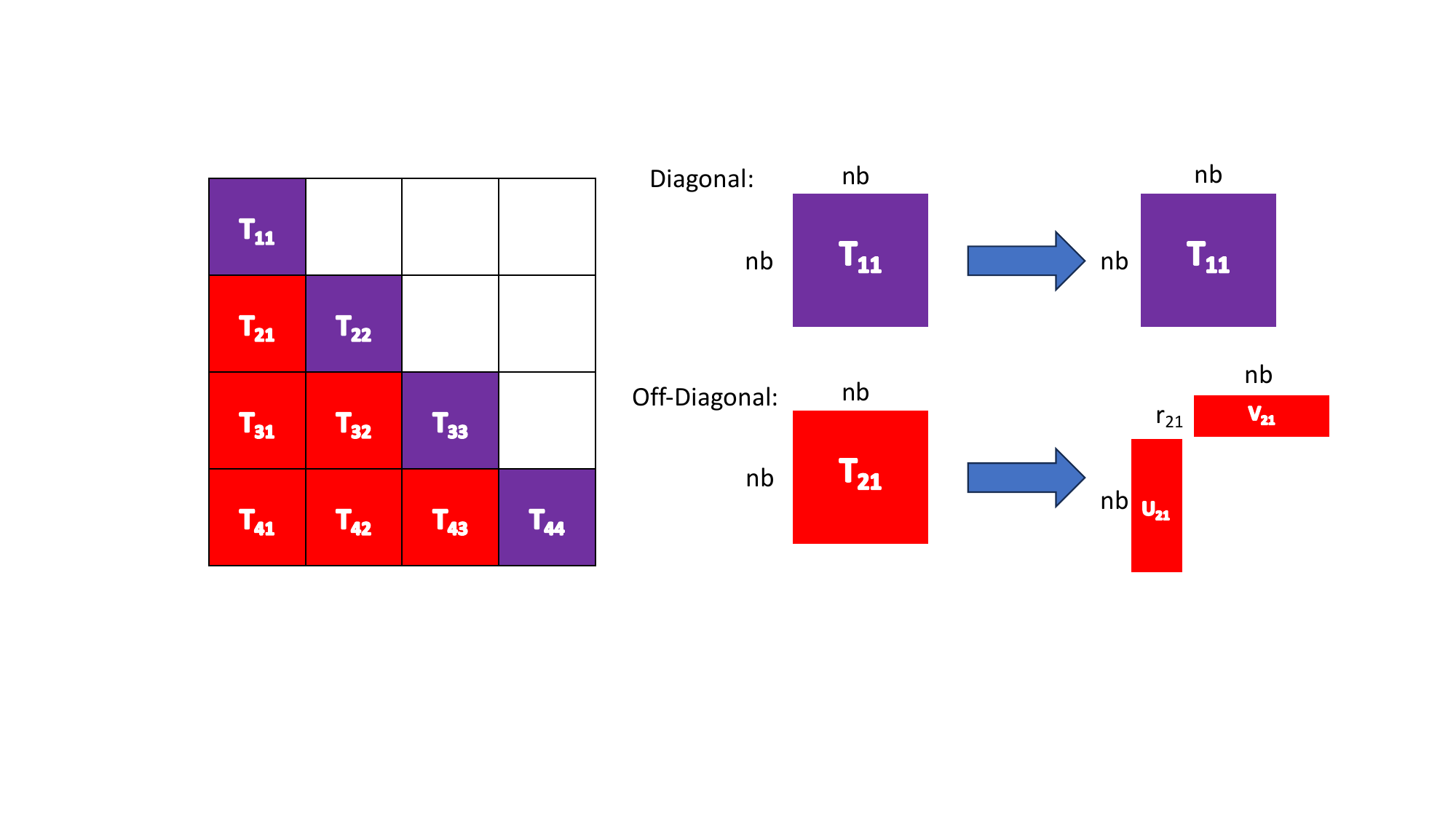}
    \vspace{-2.5cm}
    \caption{An illustration of the tile low-rank compression. The diagonal tiles are kept unchanged, while the off-diagonal tiles are compressed via singular value decomposition and then stored into smaller matrices, $\bf U$ and $\bf V$, instead of the original dense matrix tiles of size $nb\times nb$. The rank of the tile $T_{21}$ in this figure is $r_{21}$. }
    \label{tlr}
\end{figure}

TLR  approximation is based on approximating each off-diagonal tile using a low-rank approximation method. The approach in \cite{akbudak2017tile} utilizes Singular Value Decomposition (SVD) to derive two matrices, $\bf U$ and $\bf V$, representing the original tile's singular vectors. Each tile is compressed to a specific rank denoted by $r$, defining one dimension of these matrices, while the other dimension corresponds to the tile size, $nb$. Practically, the tile size not only impacts the accuracy of the compression but also plays a critical role in the performance of linear solvers during runtime. Figure~\ref{tlr} illustrates an example of compressing a dense tile into the two matrices $\bf U$ and $\bf V$.

Rank distribution also depends on how the coordinates in the covariance matrix are ordered. The ordering strategy can significantly influence the effectiveness of the TLR approximation, as it determines the structure and pattern of interactions within the matrix. Different ordering techniques can lead to varying compression efficiencies and ranks in the resulting tiles, thereby impacting the overall performance and accuracy of the TLR approximation. It is crucial to choose an appropriate ordering method that aligns with the specific characteristics and requirements of the data and the computational objectives at hand. In the subsequent section, we summarize various ordering algorithms from the literature studied in this work.


\section{Spatial Ordering Methods}
\label{section:ordering}

Spatial ordering is crucial for ordering spatial data in many fields, including spatial statistics, computer science, and geography, to efficiently organize and manage spatial data. The main goal is to transform the spatial relations into a one-dimensional structure while preserving the spatial locality. In one-dimensional coordinate systems, such as those used for time series data, a natural order is dictated by the progression along the real line. This inherent ordering means that adjacent data points in a time series are typically more correlated, following a sequential arrangement based on time. However, this natural ordering does not exist in multi-dimensional coordinate systems. In these systems, coordinates extend across multiple dimensions and lack a straightforward, inherent sequence like their one-dimensional counterparts. Assuming creating a covariance matrix based on these data, large values will be mostly located near the diagonal, and the off-diagonal elements will tend to become closer to zero. However, for spatial data with multi-dimensional coordinates, such property is not granted if we order the locations randomly, and there is a risk of having many large values in the off-diagonal part of the covariance matrix. Therefore, choosing the order of the locations wisely can be a crucial step in optimizing the computation.

The TLR approximation operates based on a user-defined accuracy threshold, which dictates how many singular values are used for off-diagonal tile compression. TLR assumes that diagonal tiles are kept in a dense format. A smaller number of singular values results in more compressed tiles, albeit with a greater loss of information. This chosen number of singular values is termed ``rank'', denoted by $r$. Reducing $r$ compresses the matrix, balancing compression and information retention. For the TLR approximation to be effective in terms of memory and computation, assuming that all tiles in the matrix are square and of size $nb \times nb$, the rank $r$ should ideally be less than $2\times(nb/2)$, to ensure that the number of elements in both $\mathbf{U}$ and $\mathbf{V}$ are less than $(nb \times nb)/2$.

Given a specific user-defined threshold, the arrangement of elements in the covariance matrix significantly impacts the ranks of off-diagonal tiles when using TLR approximation. Typically, there are two methods to decrease the ranks of the off-diagonal tiles:

\begin{itemize}

    \item To reorder the $n$-dimensional locations before generating the covariance matrix. The goal is to ensure that locations adjacent in the final one-dimensional order are also neighboring in the original multi-dimensional space. As a result, the larger values in the covariance matrix will be clustered around the diagonal, leading to lower ranks for the off-diagonal tiles.

    \item To directly reorganize the covariance matrix based on the value of each entry. This approach will result in lower ranks for the off-diagonal tiles of the covariance matrix.
\end{itemize}

In this study, we integrated various ordering algorithms into the \textit{ExaGeoStat} software to examine their effects on the accuracy and performance of the TLR approximation. Our evaluation is based on a range of spatial statistics covariance functions. The subsequent subsections provide a concise overview of the different ordering methods employed.

\subsection{Space-Filling Curves}
Space-filling curves are unique in their ability to occupy any given space using a one-dimensional line, regardless of its dimension. This concept has been utilized for decades to transform points in high-dimensional spaces into a one-dimensional arrangement. Here, we consider a set of randomly selected points within a unit hypercube of dimension $d$, where $d \in \mathbb{Z}^+$. By applying a space-filling curve to encompass this unit hypercube, each point becomes associated with a specific position on the curve. Now, envision straightening this curve into a one-dimensional line. As the curve transforms, the points align along this line, establishing a sequential order. The Morton and Hilbert curves are notable examples of space-filling curves used for this ordering process.

\subsubsection{Morton Curve}
The Morton curve (\citealp{morton}), or the Z-order curve, is a curve that covers all integers in the interior of the $d$-dimensional hypercube $[0, 2^p)^d$ of size $2^p$. The curve is constructed by interleaving the binary representation of the coordinates of each point in the $d$-dimensional space. Each point is converted to a single scalar value and can be easily sorted using traditional algorithms.

\begin{figure}[h!]
    \centering
    \includegraphics[width=0.9\linewidth]{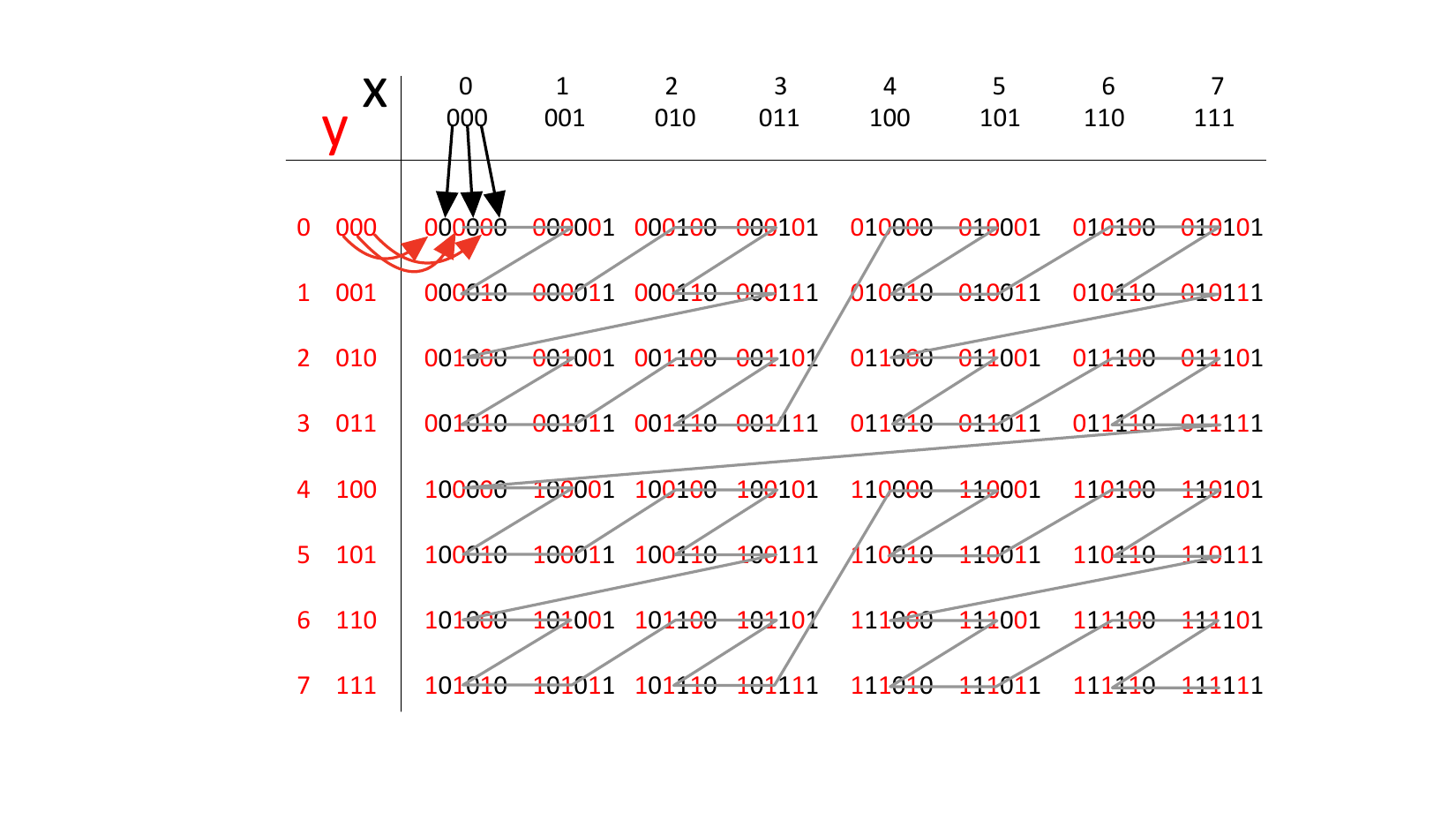}
    \vspace{-.7cm}
    \caption{An illustration of the $2$-d Morton curve that covers the interior of the two-dimensional hypercube $[0, 2^p)^2$, with $p=3$, as well as its encoding with binary numbers. In our implementation, for each $2$-d location, we find its corresponding $1$-d index on the Morton curve by interleaving the binary digits of its $2$-d coordinates, as shown above.}
    \label{morton}
\end{figure}

An illustration of the Morton curve in the two-dimensional case is shown in Figure \ref{morton}, which is a recursively Z-shaped curve linking the integer-valued coordinates in the two-dimensional plane. Higher-dimensional Morton curves are designed similarly.

As shown in Figure \ref{morton}, it is evident that most adjacent points on the original two-dimensional grid are similarly positioned nearby along the Morton curve. This alignment underscores the objective of maintaining spatial ordering in the transformation process. Figure \ref{morton} also illustrates that the `Z' patterns divide the space into $2\times2$ grids in the two-dimensional scenario. For any given location in this two-dimensional space, its corresponding one-dimensional index on the Morton curve can be determined by interleaving the binary values of its coordinates. This technique is adaptable for implementing Morton ordering in various dimensional spaces.

\subsubsection{Hilbert Curve}
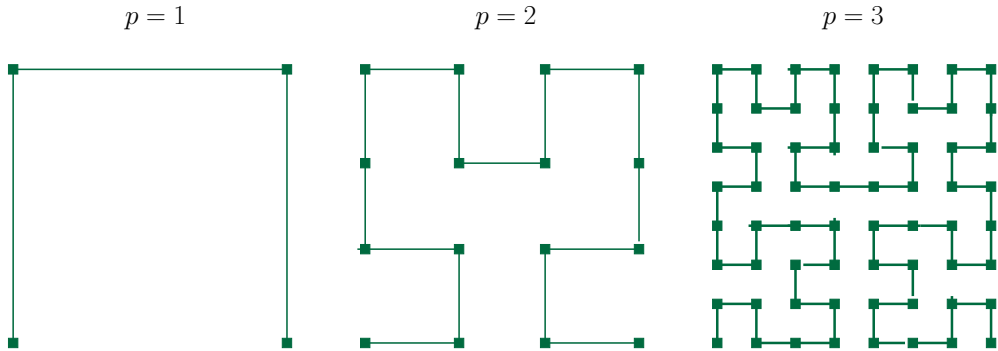
\begin{figure}[h!]
\centering
\resizebox{0.8\textwidth}{!}{
\begin{circuitikz}
\tikzstyle{every node}=[font=\small]
\node[label={\Huge \bf $p=1$}] (a) at (5.8, 14){};
\node[label={\Huge \bf $p=2$}] (a) at (17, 14){};
\node[label={\Huge \bf $p=3$}] (a) at (28.1, 14){};
\draw [ color={rgb,255:red,0;green,106;blue,63}, fill={rgb,255:red,0; green,107; blue,63} ] (1.1,13.15) rectangle (1.4,12.85);
\draw [ color={rgb,255:red,0;green,106;blue,63}, fill={rgb,255:red,0; green,107; blue,63} , line width=0.02pt ] (1.1,4.4) rectangle (1.4,4.1);
\draw [ color={rgb,255:red,0;green,106;blue,63}, fill={rgb,255:red,0; green,107; blue,63} , line width=0.02pt ] (9.85,13.15) rectangle (10.15,12.85);
\draw [ color={rgb,255:red,0;green,106;blue,63}, fill={rgb,255:red,0; green,107; blue,63} , line width=0.02pt ] (9.85,4.4) rectangle (10.15,4.1);
\draw [color = {rgb,255:red,0;green,106; blue,63}, line width=1.4pt, short] (1.25,13) .. controls (1.25,8.75) and (1.25,8.75) .. (1.25,4.25);
\draw [color = {rgb,255:red,0;green,106; blue,63}, line width=1.4pt, short] (1.25,13) .. controls (5.75,13) and (5.75,13) .. (10,13);
\draw [color = {rgb,255:red,0;green,106; blue,63}, line width=1.4pt, short] (10,13) .. controls (10,8.75) and (10,8.75) .. (10,4.25);
\draw [ color={rgb,255:red,0;green,106;blue,63}, fill={rgb,255:red,0; green,107; blue,63} , line width=0.02pt ] (12.35,4.4) rectangle (12.65,4.1);
\draw [ color={rgb,255:red,0;green,106;blue,63}, fill={rgb,255:red,0; green,107; blue,63} , line width=0.02pt ] (21.1,4.4) rectangle (21.4,4.1);
\draw [ color={rgb,255:red,0;green,106;blue,63}, fill={rgb,255:red,0; green,107; blue,63} , line width=0.02pt ] (15.35,4.4) rectangle (15.65,4.1);
\draw [ color={rgb,255:red,0;green,106;blue,63}, fill={rgb,255:red,0; green,107; blue,63} , line width=0.02pt ] (18.1,4.4) rectangle (18.4,4.1);
\draw [ color={rgb,255:red,0;green,106;blue,63}, fill={rgb,255:red,0; green,107; blue,63} , line width=0.02pt ] (12.35,13.15) rectangle (12.65,12.85);
\draw [ color={rgb,255:red,0;green,106;blue,63}, fill={rgb,255:red,0; green,107; blue,63} , line width=0.02pt ] (15.35,13.15) rectangle (15.65,12.85);
\draw [ color={rgb,255:red,0;green,106;blue,63}, fill={rgb,255:red,0; green,107; blue,63} , line width=0.02pt ] (18.1,13.15) rectangle (18.4,12.85);
\draw [ color={rgb,255:red,0;green,106;blue,63}, fill={rgb,255:red,0; green,107; blue,63} , line width=0.02pt ] (21.1,13.15) rectangle (21.4,12.85);
\draw [ color={rgb,255:red,0;green,106;blue,63}, fill={rgb,255:red,0; green,107; blue,63} , line width=0.02pt ] (12.35,10.15) rectangle (12.65,9.85);
\draw [ color={rgb,255:red,0;green,106;blue,63}, fill={rgb,255:red,0; green,107; blue,63} , line width=0.02pt ] (12.35,7.4) rectangle (12.65,7.1);
\draw [ color={rgb,255:red,0;green,106;blue,63}, fill={rgb,255:red,0; green,107; blue,63} , line width=0.02pt ] (15.35,10.15) rectangle (15.65,9.85);
\draw [ color={rgb,255:red,0;green,106;blue,63}, fill={rgb,255:red,0; green,107; blue,63} , line width=0.02pt ] (15.35,7.4) rectangle (15.65,7.1);
\draw [ color={rgb,255:red,0;green,106;blue,63}, fill={rgb,255:red,0; green,107; blue,63} , line width=0.02pt ] (18.1,10.15) rectangle (18.4,9.85);
\draw [ color={rgb,255:red,0;green,106;blue,63}, fill={rgb,255:red,0; green,107; blue,63} , line width=0.02pt ] (18.1,7.4) rectangle (18.4,7.1);
\draw [ color={rgb,255:red,0;green,106;blue,63}, fill={rgb,255:red,0; green,107; blue,63} , line width=0.02pt ] (21.1,10.15) rectangle (21.4,9.85);
\draw [ color={rgb,255:red,0;green,106;blue,63}, fill={rgb,255:red,0; green,107; blue,63} , line width=0.02pt ] (21.1,7.4) rectangle (21.4,7.1);
\draw [color = {rgb,255:red,0;green,106; blue,63}, line width=1.4pt, short] (12.5,4.25) .. controls (14,4.25) and (14,4.25) .. (15.5,4.25);
\draw [color = {rgb,255:red,0;green,106; blue,63}, line width=1.4pt, short] (15.5,4.25) .. controls (15.5,5.75) and (15.5,5.75) .. (15.5,7.25);
\draw [color = {rgb,255:red,0;green,106; blue,63}, line width=1.4pt, short] (15.5,7.25) .. controls (14,7.25) and (14,7.25) .. (12.25,7.25);
\draw [color = {rgb,255:red,0;green,106; blue,63}, line width=1.4pt, short] (12.5,7.25) .. controls (12.5,8.75) and (12.5,8.75) .. (12.5,10);
\draw [color = {rgb,255:red,0;green,106; blue,63}, line width=1.4pt, short] (12.5,10) .. controls (12.5,11.5) and (12.5,11.5) .. (12.5,13);
\draw [color = {rgb,255:red,0;green,106; blue,63}, line width=1.4pt, short] (12.5,13) .. controls (14,13) and (14,13) .. (15.5,13);
\draw [color = {rgb,255:red,0;green,106; blue,63}, line width=1.4pt, short] (15.5,13) .. controls (15.5,11.5) and (15.5,11.5) .. (15.5,10);
\draw [color = {rgb,255:red,0;green,106; blue,63}, line width=1.4pt, short] (15.5,10) .. controls (17,10) and (17,10) .. (18.25,10);
\draw [color = {rgb,255:red,0;green,106; blue,63}, line width=1.4pt, short] (18.25,10) .. controls (18.25,11.5) and (18.25,11.5) .. (18.25,13);
\draw [color = {rgb,255:red,0;green,106; blue,63}, line width=1.4pt, short] (18.25,13) .. controls (19.75,13) and (19.75,13) .. (21.25,13);
\draw [color = {rgb,255:red,0;green,106; blue,63}, line width=1.4pt, short] (21.25,13) .. controls (21.25,11.5) and (21.25,11.5) .. (21.25,10);
\draw [color = {rgb,255:red,0;green,106; blue,63}, line width=1.4pt, short] (21.25,10) .. controls (21.25,8.75) and (21.25,8.75) .. (21.25,7.5);
\draw [color = {rgb,255:red,0;green,106; blue,63}, line width=1.4pt, short] (21.25,7.25) .. controls (19.75,7.25) and (19.75,7.25) .. (18.25,7.25);
\draw [color = {rgb,255:red,0;green,106; blue,63}, line width=1.4pt, short] (18.25,7.25) .. controls (18.25,5.75) and (18.25,5.75) .. (18.25,4.25);
\draw [color = {rgb,255:red,0;green,106; blue,63}, line width=1.4pt, short] (18.25,4.25) .. controls (19.75,4.25) and (19.75,4.25) .. (21.25,4.25);
\draw [ color={rgb,255:red,0;green,106;blue,63}, fill={rgb,255:red,0; green,107; blue,63} , line width=0.02pt ] (23.6,13.15) rectangle (23.9,12.85);
\draw [ color={rgb,255:red,0;green,106;blue,63}, fill={rgb,255:red,0; green,107; blue,63} , line width=0.02pt ] (24.85,13.15) rectangle (25.15,12.85);
\draw [ color={rgb,255:red,0;green,106;blue,63}, fill={rgb,255:red,0; green,107; blue,63} , line width=0.02pt ] (26.1,13.15) rectangle (26.4,12.85);
\draw [ color={rgb,255:red,0;green,106;blue,63}, fill={rgb,255:red,0; green,107; blue,63} , line width=0.02pt ] (27.35,13.15) rectangle (27.65,12.85);
\draw [ color={rgb,255:red,0;green,106;blue,63}, fill={rgb,255:red,0; green,107; blue,63} , line width=0.02pt ] (28.6,13.15) rectangle (28.9,12.85);
\draw [ color={rgb,255:red,0;green,106;blue,63}, fill={rgb,255:red,0; green,107; blue,63} , line width=0.02pt ] (29.85,13.15) rectangle (30.15,12.85);
\draw [ color={rgb,255:red,0;green,106;blue,63}, fill={rgb,255:red,0; green,107; blue,63} , line width=0.02pt ] (31.1,13.15) rectangle (31.4,12.85);
\draw [ color={rgb,255:red,0;green,106;blue,63}, fill={rgb,255:red,0; green,107; blue,63} , line width=0.02pt ] (23.6,11.9) rectangle (23.9,11.6);
\draw [ color={rgb,255:red,0;green,106;blue,63}, fill={rgb,255:red,0; green,107; blue,63} , line width=0.02pt ] (24.85,11.9) rectangle (25.15,11.6);
\draw [ color={rgb,255:red,0;green,106;blue,63}, fill={rgb,255:red,0; green,107; blue,63} , line width=0.02pt ] (26.1,11.9) rectangle (26.4,11.6);
\draw [ color={rgb,255:red,0;green,106;blue,63}, fill={rgb,255:red,0; green,107; blue,63} , line width=0.02pt ] (27.35,11.9) rectangle (27.65,11.6);
\draw [ color={rgb,255:red,0;green,106;blue,63}, fill={rgb,255:red,0; green,107; blue,63} , line width=0.02pt ] (28.6,11.9) rectangle (28.9,11.6);
\draw [ color={rgb,255:red,0;green,106;blue,63}, fill={rgb,255:red,0; green,107; blue,63} , line width=0.02pt ] (29.85,11.9) rectangle (30.15,11.6);
\draw [ color={rgb,255:red,0;green,106;blue,63}, fill={rgb,255:red,0; green,107; blue,63} , line width=0.02pt ] (31.1,11.9) rectangle (31.4,11.6);
\draw [ color={rgb,255:red,0;green,106;blue,63}, fill={rgb,255:red,0; green,107; blue,63} , line width=0.02pt ] (23.6,10.65) rectangle (23.9,10.35);
\draw [ color={rgb,255:red,0;green,106;blue,63}, fill={rgb,255:red,0; green,107; blue,63} , line width=0.02pt ] (24.85,10.65) rectangle (25.15,10.35);
\draw [ color={rgb,255:red,0;green,106;blue,63}, fill={rgb,255:red,0; green,107; blue,63} , line width=0.02pt ] (26.1,10.65) rectangle (26.4,10.35);
\draw [ color={rgb,255:red,0;green,106;blue,63}, fill={rgb,255:red,0; green,107; blue,63} , line width=0.02pt ] (27.35,10.65) rectangle (27.65,10.35);
\draw [ color={rgb,255:red,0;green,106;blue,63}, fill={rgb,255:red,0; green,107; blue,63} , line width=0.02pt ] (28.6,10.65) rectangle (28.9,10.35);
\draw [ color={rgb,255:red,0;green,106;blue,63}, fill={rgb,255:red,0; green,107; blue,63} , line width=0.02pt ] (29.85,10.65) rectangle (30.15,10.35);
\draw [ color={rgb,255:red,0;green,106;blue,63}, fill={rgb,255:red,0; green,107; blue,63} , line width=0.02pt ] (31.1,10.65) rectangle (31.4,10.35);
\draw [ color={rgb,255:red,0;green,106;blue,63}, fill={rgb,255:red,0; green,107; blue,63} , line width=0.02pt ] (23.6,9.4) rectangle (23.9,9.1);
\draw [ color={rgb,255:red,0;green,106;blue,63}, fill={rgb,255:red,0; green,107; blue,63} , line width=0.02pt ] (24.85,9.4) rectangle (25.15,9.1);
\draw [ color={rgb,255:red,0;green,106;blue,63}, fill={rgb,255:red,0; green,107; blue,63} , line width=0.02pt ] (26.1,9.4) rectangle (26.4,9.1);
\draw [ color={rgb,255:red,0;green,106;blue,63}, fill={rgb,255:red,0; green,107; blue,63} , line width=0.02pt ] (28.6,9.4) rectangle (28.9,9.1);
\draw [ color={rgb,255:red,0;green,106;blue,63}, fill={rgb,255:red,0; green,107; blue,63} , line width=0.02pt ] (29.85,9.4) rectangle (30.15,9.1);
\draw [ color={rgb,255:red,0;green,106;blue,63}, fill={rgb,255:red,0; green,107; blue,63} , line width=0.02pt ] (31.1,9.4) rectangle (31.4,9.1);
\draw [ color={rgb,255:red,0;green,106;blue,63}, fill={rgb,255:red,0; green,107; blue,63} , line width=0.02pt ] (27.35,9.4) rectangle (27.65,9.1);
\draw [ color={rgb,255:red,0;green,106;blue,63}, fill={rgb,255:red,0; green,107; blue,63} , line width=0.02pt ] (23.6,8.15) rectangle (23.9,7.85);
\draw [ color={rgb,255:red,0;green,106;blue,63}, fill={rgb,255:red,0; green,107; blue,63} , line width=0.02pt ] (24.85,8.15) rectangle (25.15,7.85);
\draw [ color={rgb,255:red,0;green,106;blue,63}, fill={rgb,255:red,0; green,107; blue,63} , line width=0.02pt ] (26.1,8.15) rectangle (26.4,7.85);
\draw [ color={rgb,255:red,0;green,106;blue,63}, fill={rgb,255:red,0; green,107; blue,63} , line width=0.02pt ] (27.35,8.15) rectangle (27.65,7.85);
\draw [ color={rgb,255:red,0;green,106;blue,63}, fill={rgb,255:red,0; green,107; blue,63} , line width=0.02pt ] (28.6,8.15) rectangle (28.9,7.85);
\draw [ color={rgb,255:red,0;green,106;blue,63}, fill={rgb,255:red,0; green,107; blue,63} , line width=0.02pt ] (29.85,8.15) rectangle (30.15,7.85);
\draw [ color={rgb,255:red,0;green,106;blue,63}, fill={rgb,255:red,0; green,107; blue,63} , line width=0.02pt ] (31.1,8.15) rectangle (31.4,7.85);
\draw [ color={rgb,255:red,0;green,106;blue,63}, fill={rgb,255:red,0; green,107; blue,63} , line width=0.02pt ] (23.6,6.9) rectangle (23.9,6.6);
\draw [ color={rgb,255:red,0;green,106;blue,63}, fill={rgb,255:red,0; green,107; blue,63} , line width=0.02pt ] (24.85,6.9) rectangle (25.15,6.6);
\draw [ color={rgb,255:red,0;green,106;blue,63}, fill={rgb,255:red,0; green,107; blue,63} , line width=0.02pt ] (26.1,6.9) rectangle (26.4,6.6);
\draw [ color={rgb,255:red,0;green,106;blue,63}, fill={rgb,255:red,0; green,107; blue,63} , line width=0.02pt ] (27.35,6.9) rectangle (27.65,6.6);
\draw [ color={rgb,255:red,0;green,106;blue,63}, fill={rgb,255:red,0; green,107; blue,63} , line width=0.02pt ] (28.6,6.9) rectangle (28.9,6.6);
\draw [ color={rgb,255:red,0;green,106;blue,63}, fill={rgb,255:red,0; green,107; blue,63} , line width=0.02pt ] (29.85,6.9) rectangle (30.15,6.6);
\draw [ color={rgb,255:red,0;green,106;blue,63}, fill={rgb,255:red,0; green,107; blue,63} , line width=0.02pt ] (31.1,6.9) rectangle (31.4,6.6);
\draw [ color={rgb,255:red,0;green,106;blue,63}, fill={rgb,255:red,0; green,107; blue,63} , line width=0.02pt ] (23.6,5.65) rectangle (23.9,5.35);
\draw [ color={rgb,255:red,0;green,106;blue,63}, fill={rgb,255:red,0; green,107; blue,63} , line width=0.02pt ] (24.85,5.65) rectangle (25.15,5.35);
\draw [ color={rgb,255:red,0;green,106;blue,63}, fill={rgb,255:red,0; green,107; blue,63} , line width=0.02pt ] (26.1,5.65) rectangle (26.4,5.35);
\draw [ color={rgb,255:red,0;green,106;blue,63}, fill={rgb,255:red,0; green,107; blue,63} , line width=0.02pt ] (27.35,5.65) rectangle (27.65,5.35);
\draw [ color={rgb,255:red,0;green,106;blue,63}, fill={rgb,255:red,0; green,107; blue,63} , line width=0.02pt ] (28.6,5.65) rectangle (28.9,5.35);
\draw [ color={rgb,255:red,0;green,106;blue,63}, fill={rgb,255:red,0; green,107; blue,63} , line width=0.7pt ] (29.85,5.65) rectangle (30.15,5.35);
\draw [ color={rgb,255:red,0;green,106;blue,63}, fill={rgb,255:red,0; green,107; blue,63} , line width=0.02pt ] (31.1,5.65) rectangle (31.4,5.35);
\draw [ color={rgb,255:red,0;green,106;blue,63}, fill={rgb,255:red,0; green,107; blue,63} , line width=0.02pt ] (23.6,4.4) rectangle (23.9,4.1);
\draw [ color={rgb,255:red,0;green,106;blue,63}, fill={rgb,255:red,0; green,107; blue,63} , line width=0.02pt ] (24.85,4.4) rectangle (25.15,4.1);
\draw [ color={rgb,255:red,0;green,106;blue,63}, fill={rgb,255:red,0; green,107; blue,63} , line width=0.02pt ] (26.1,4.4) rectangle (26.4,4.1);
\draw [ color={rgb,255:red,0;green,106;blue,63}, fill={rgb,255:red,0; green,107; blue,63} , line width=0.02pt ] (27.35,4.4) rectangle (27.65,4.1);
\draw [ color={rgb,255:red,0;green,106;blue,63}, fill={rgb,255:red,0; green,107; blue,63} , line width=0.02pt ] (28.6,4.4) rectangle (28.9,4.1);
\draw [ color={rgb,255:red,0;green,106;blue,63}, fill={rgb,255:red,0; green,107; blue,63} , line width=0.02pt ] (29.85,4.4) rectangle (30.15,4.1);
\draw [ color={rgb,255:red,0;green,106;blue,63}, fill={rgb,255:red,0; green,107; blue,63} , line width=0.02pt ] (31.1,4.4) rectangle (31.4,4.1);
\draw [color={rgb,255:red,0;green,107;blue,63}, line width=2.3pt, short] (23.75,4.25) .. controls (23.75,5) and (23.75,5) .. (23.75,5.5);
\draw [color={rgb,255:red,0;green,107;blue,63}, line width=2.3pt, short] (23.75,5.5) .. controls (24.5,5.5) and (24.5,5.5) .. (25,5.5);
\draw [color={rgb,255:red,0;green,107;blue,63}, line width=2.3pt, short] (25,5.5) .. controls (25,5) and (25,5) .. (25,4.25);
\draw [color={rgb,255:red,0;green,107;blue,63}, line width=2.3pt, short] (25,4.25) .. controls (25.75,4.25) and (25.75,4.25) .. (26.25,4.25);
\draw [color={rgb,255:red,0;green,107;blue,63}, line width=2.3pt, short] (26.25,4.25) .. controls (27,4.25) and (27,4.25) .. (27.5,4.25);
\draw [ color={rgb,255:red,0;green,106;blue,63}, fill={rgb,255:red,0; green,107; blue,63} , line width=0.02pt ] (32.35,13.15) rectangle (32.65,12.85);
\draw [ color={rgb,255:red,0;green,106;blue,63}, fill={rgb,255:red,0; green,107; blue,63} , line width=0.02pt ] (32.35,11.9) rectangle (32.65,11.6);
\draw [ color={rgb,255:red,0;green,106;blue,63}, fill={rgb,255:red,0; green,107; blue,63} , line width=0.02pt ] (32.35,10.65) rectangle (32.65,10.35);
\draw [ color={rgb,255:red,0;green,106;blue,63}, fill={rgb,255:red,0; green,107; blue,63} , line width=0.02pt ] (32.35,9.4) rectangle (32.65,9.1);
\draw [ color={rgb,255:red,0;green,106;blue,63}, fill={rgb,255:red,0; green,107; blue,63} , line width=0.02pt ] (32.35,8.15) rectangle (32.65,7.85);
\draw [ color={rgb,255:red,0;green,106;blue,63}, fill={rgb,255:red,0; green,107; blue,63} , line width=0.02pt ] (32.35,6.9) rectangle (32.65,6.6);
\draw [ color={rgb,255:red,0;green,106;blue,63}, fill={rgb,255:red,0; green,107; blue,63} , line width=0.02pt ] (32.35,5.65) rectangle (32.65,5.35);
\draw [ color={rgb,255:red,0;green,106;blue,63}, fill={rgb,255:red,0; green,107; blue,63} , line width=0.02pt ] (32.35,4.4) rectangle (32.65,4.1);
\draw [color={rgb,255:red,0;green,107;blue,63}, line width=2.3pt, short] (27.5,4.25) .. controls (27.5,5) and (27.5,5) .. (27.5,5.5);
\draw [color={rgb,255:red,0;green,107;blue,63}, line width=2.3pt, short] (27.5,5.5) .. controls (27,5.5) and (27,5.5) .. (26.25,5.5);
\draw [color={rgb,255:red,0;green,107;blue,63}, line width=2.3pt, short] (26.25,5.5) .. controls (26.25,6.25) and (26.25,6.25) .. (26.25,6.75);
\draw [color={rgb,255:red,0;green,107;blue,63}, line width=2.3pt, short] (26.5,6.75) .. controls (27,6.75) and (27,6.75) .. (27.5,6.75);
\draw [color={rgb,255:red,0;green,107;blue,63}, line width=2.3pt, short] (27.5,6.75) .. controls (27.5,7.5) and (27.5,7.5) .. (27.5,8.25);
\draw [color={rgb,255:red,0;green,107;blue,63}, line width=2.3pt, short] (27.5,8) .. controls (26.75,8) and (26.75,8) .. (26,8);
\draw [color={rgb,255:red,0;green,107;blue,63}, line width=2.3pt, short] (26.25,8) .. controls (25.5,8) and (25.5,8) .. (24.75,8);
\draw [color={rgb,255:red,0;green,107;blue,63}, line width=2.3pt, short] (25,8) .. controls (25,7.5) and (25,7.5) .. (25,6.75);
\draw [color={rgb,255:red,0;green,107;blue,63}, line width=2.3pt, short] (25,6.75) .. controls (24.5,6.75) and (24.5,6.75) .. (23.75,6.75);
\draw [color={rgb,255:red,0;green,107;blue,63}, line width=2.3pt, short] (23.75,6.75) .. controls (23.75,7.5) and (23.75,7.5) .. (23.75,8);
\draw [color={rgb,255:red,0;green,107;blue,63}, line width=2.3pt, short] (23.75,8) .. controls (23.75,8.75) and (23.75,8.75) .. (23.75,9.25);
\draw [color={rgb,255:red,0;green,107;blue,63}, line width=2.3pt, short] (23.75,9.25) .. controls (24.5,9.25) and (24.5,9.25) .. (25,9.25);
\draw [color={rgb,255:red,0;green,107;blue,63}, line width=2.3pt, short] (25,9.25) .. controls (25,10) and (25,10) .. (25,10.5);
\draw [color={rgb,255:red,0;green,107;blue,63}, line width=2.3pt, short] (25,10.5) .. controls (24.5,10.5) and (24.5,10.5) .. (23.75,10.5);
\draw [color={rgb,255:red,0;green,107;blue,63}, line width=2.3pt, short] (23.75,10.5) .. controls (23.75,11.25) and (23.75,11.25) .. (23.75,11.75);
\draw [color={rgb,255:red,0;green,107;blue,63}, line width=2.3pt, short] (23.75,11.75) .. controls (23.75,12.5) and (23.75,12.5) .. (23.75,13);
\draw [color={rgb,255:red,0;green,107;blue,63}, line width=2.3pt, short] (23.75,13) .. controls (24.5,13) and (24.5,13) .. (25,13);
\draw [color={rgb,255:red,0;green,107;blue,63}, line width=2.3pt, short] (25,13) .. controls (25,12.5) and (25,12.5) .. (25,11.75);
\draw [color={rgb,255:red,0;green,107;blue,63}, line width=2.3pt, short] (25,11.75) .. controls (25.75,11.75) and (25.75,11.75) .. (26.25,11.75);
\draw [color={rgb,255:red,0;green,107;blue,63}, line width=2.3pt, short] (26.25,11.75) .. controls (26.25,12.5) and (26.25,12.5) .. (26.25,13);
\draw [color={rgb,255:red,0;green,107;blue,63}, line width=2.3pt, short] (26,13) .. controls (26.75,13) and (26.75,13) .. (27.5,13);
\draw [color={rgb,255:red,0;green,107;blue,63}, line width=2.3pt, short] (27.5,13) .. controls (27.5,12.5) and (27.5,12.5) .. (27.5,11.75);
\draw [color={rgb,255:red,0;green,107;blue,63}, line width=2.3pt, short] (27.5,11.75) .. controls (27.5,11) and (27.5,11) .. (27.5,10.25);
\draw [color={rgb,255:red,0;green,107;blue,63}, line width=2.3pt, short] (27.5,10.5) .. controls (26.75,10.5) and (26.75,10.5) .. (26,10.5);
\draw [color={rgb,255:red,0;green,107;blue,63}, line width=2.3pt, short] (26.25,10.5) .. controls (26.25,10) and (26.25,10) .. (26.25,9.25);
\draw [color={rgb,255:red,0;green,107;blue,63}, line width=2.3pt, short] (26.25,9.25) .. controls (27,9.25) and (27,9.25) .. (27.5,9.25);
\draw [color={rgb,255:red,0;green,107;blue,63}, line width=2.3pt, short] (27.5,9.25) .. controls (28.25,9.25) and (28.25,9.25) .. (28.75,9.25);
\draw [color={rgb,255:red,0;green,107;blue,63}, line width=2.3pt, short] (28.75,9.25) .. controls (29.5,9.25) and (29.5,9.25) .. (30,9.25);
\draw [color={rgb,255:red,0;green,107;blue,63}, line width=2.3pt, short] (30,9.25) .. controls (30,10) and (30,10) .. (30,10.5);
\draw [color={rgb,255:red,0;green,107;blue,63}, line width=2.3pt, short] (30,10.5) .. controls (29.5,10.5) and (29.5,10.5) .. (29,10.5);
\draw [color={rgb,255:red,0;green,107;blue,63}, line width=2.3pt, short] (28.75,10.5) .. controls (28.75,11.25) and (28.75,11.25) .. (28.75,11.75);
\draw [color={rgb,255:red,0;green,107;blue,63}, line width=2.3pt, short] (28.75,11.75) .. controls (28.75,12.5) and (28.75,12.5) .. (28.75,13);
\draw [color={rgb,255:red,0;green,107;blue,63}, line width=2.3pt, short] (28.75,13) .. controls (29.5,13) and (29.5,13) .. (30,13);
\draw [color={rgb,255:red,0;green,107;blue,63}, line width=2.3pt, short] (30,13) .. controls (30,12.5) and (30,12.5) .. (30,12);
\draw [color={rgb,255:red,0;green,107;blue,63}, line width=2.3pt, short] (30,11.75) .. controls (30.75,11.75) and (30.75,11.75) .. (31.25,11.75);
\draw [color={rgb,255:red,0;green,107;blue,63}, line width=2.3pt, short] (31.25,11.75) .. controls (31.25,12.5) and (31.25,12.5) .. (31.25,13);
\draw [color={rgb,255:red,0;green,107;blue,63}, line width=2.3pt, short] (31.25,13) .. controls (32,13) and (32,13) .. (32.5,13);
\draw [color={rgb,255:red,0;green,107;blue,63}, line width=2.3pt, short] (32.5,13) .. controls (32.5,12.25) and (32.5,12.25) .. (32.5,11.5);
\draw [color={rgb,255:red,0;green,107;blue,63}, line width=2.3pt, short] (32.5,11.75) .. controls (32.5,11.25) and (32.5,11.25) .. (32.5,10.5);
\draw [color={rgb,255:red,0;green,107;blue,63}, line width=2.3pt, short] (32.5,10.5) .. controls (32,10.5) and (32,10.5) .. (31.25,10.5);
\draw [color={rgb,255:red,0;green,107;blue,63}, line width=2.3pt, short] (31.25,10.5) .. controls (31.25,10) and (31.25,10) .. (31.25,9.25);
\draw [color={rgb,255:red,0;green,107;blue,63}, line width=2.3pt, short] (31.25,9.25) .. controls (32,9.25) and (32,9.25) .. (32.5,9.25);
\draw [color={rgb,255:red,0;green,107;blue,63}, line width=2.3pt, short] (32.5,9.25) .. controls (32.5,8.75) and (32.5,8.75) .. (32.5,8);
\draw [color={rgb,255:red,0;green,107;blue,63}, line width=2.3pt, short] (32.5,8) .. controls (32.5,7.5) and (32.5,7.5) .. (32.5,6.75);
\draw [color={rgb,255:red,0;green,107;blue,63}, line width=2.3pt, short] (32.5,6.75) .. controls (32,6.75) and (32,6.75) .. (31.25,6.75);
\draw [color={rgb,255:red,0;green,107;blue,63}, line width=2.3pt, short] (31.25,6.75) .. controls (31.25,7.5) and (31.25,7.5) .. (31.25,8);
\draw [color={rgb,255:red,0;green,107;blue,63}, line width=2.3pt, short] (31.25,8) .. controls (30.75,8) and (30.75,8) .. (30.25,8);
\draw [color={rgb,255:red,0;green,107;blue,63}, line width=2.3pt, short] (30.25,8) .. controls (29.5,8) and (29.5,8) .. (28.75,8);
\draw [color={rgb,255:red,0;green,107;blue,63}, line width=2.3pt, short] (28.75,8) .. controls (28.75,7.5) and (28.75,7.5) .. (28.75,6.75);
\draw [color={rgb,255:red,0;green,107;blue,63}, line width=2.3pt, short] (28.75,6.75) .. controls (29.5,6.75) and (29.5,6.75) .. (30,6.75);
\draw [color={rgb,255:red,0;green,107;blue,63}, line width=2.3pt, short] (30,6.75) .. controls (30,6.25) and (30,6.25) .. (30,5.75);
\draw [color={rgb,255:red,0;green,107;blue,63}, line width=2.3pt, short] (30,5.5) .. controls (29.5,5.5) and (29.5,5.5) .. (28.75,5.5);
\draw [color={rgb,255:red,0;green,107;blue,63}, line width=2.3pt, short] (28.75,5.5) .. controls (28.75,5) and (28.75,5) .. (28.75,4.25);
\draw [color={rgb,255:red,0;green,107;blue,63}, line width=2.3pt, short] (28.75,4.25) .. controls (29.25,4.25) and (29.25,4.25) .. (29.75,4.25);
\draw [color={rgb,255:red,0;green,107;blue,63}, line width=2.3pt, short] (30,4.25) .. controls (30.75,4.25) and (30.75,4.25) .. (31.25,4.25);
\draw [color={rgb,255:red,0;green,107;blue,63}, line width=2.3pt, short] (31.25,4.25) .. controls (31.25,5) and (31.25,5) .. (31.25,5.75);
\draw [color={rgb,255:red,0;green,107;blue,63}, line width=2.3pt, short] (31.25,5.5) .. controls (32,5.5) and (32,5.5) .. (32.5,5.5);
\draw [color={rgb,255:red,0;green,107;blue,63}, line width=2.3pt, short] (32.5,5.5) .. controls (32.5,5) and (32.5,5) .. (32.5,4.25);
\end{circuitikz}
}

\caption{An illustration of the $2$-d Hilbert curve that covers the interior of the two-dimensional hypercube $[0, 2^p)^2$, for $p=1, 2, 3$ respectively, adapted from \cite{skilling}.}
    \label{hilbert}
\end{figure}

Similar to the Morton curve, the Hilbert curve (\citealp{hilbert}) is another curve that covers all integers in the interior of the $d$-dimensional hypercube $[0, 2^p)^d$ of size $2^p$. The curve starts at one corner of the space grid and snakes through each element in a specific pattern. As the order of the curve increases, it becomes more complex, filling the space more densely. Hilbert curve aims to minimize the distance between the points close to each other in the multi-dimensional space. An illustration of the Hilbert curve in a two-dimensional case is shown in Figure \ref{hilbert}.

We have integrated the Morton and Hilbert curve ordering algorithms into the \textit{ExaGeoStat} software. For two-dimensional spatial locations, our implementations involve the following steps:

 \begin{enumerate}
        \item ``Encoding'': Convert the two-dimensional coordinates into integers. Specifically, if the location coordinates are initially in single-precision floating-point format (16 bits), we convert these into 16-bit unsigned integers. This is achieved by multiplying the coordinates by the maximum value allowable for unsigned integers in the system and then rounding the result to the nearest integer;
        \item ``Sorting'': We compute their respective one-dimensional indices for each two-dimensional coordinate point on either the Morton or Hilbert curve. Following this calculation, we sort these indices using standard sorting algorithms like quicksort;
        \item ``Decoding'': Convert the $1$-d coordinates back into $2$-d coordinates, then further convert back to floating point values in the unit square by dividing them by the maximum value of unsigned integers. 
    \end{enumerate}

The time complexity for ordering $n$ spatial locations using both Morton and Hilbert curves is identical. The encoding and decoding steps each have a time complexity of $O(n)$, while the sorting step incurs a time complexity of $O(n\log n)$. Consequently, the total time complexity for Morton and Hilbert orderings is $O(n \log n)$.

Moreover, it is important to acknowledge that when employing Morton and Hilbert orderings, converting floating-point values to integers through rounding and then reconverting them back to floating points results in coordinates that do not precisely match their original values. However, as demonstrated in our experimental findings in the subsequent section, the accuracy of parameter estimation remains largely unaffected by these minor discrepancies.

\subsection{$K$-Dimensional (KD) Tree}
A $k$-dimensional tree (KD-Tree) is a binary tree data structure that stores $k$-dimensional data where $k>1$, similar to a binary search tree for $1$-dimensional data. It can also be used for other purposes, and it has been widely applied to many different areas, not only in spatial data analysis. For $n$ spatial locations, both construction and traversal of a KD-Tree are of time complexity $O(n\log n)$. Hence, the overall time complexity for KD-Tree ordering for $n$ spatial locations is $O(n\log n)$.

In a KD-Tree, each non-leaf node stores an index of dimension $i$ and a value $m$, and it functions as a hyperplane dividing the $k$-dimensional space into two half-spaces: all inputs with value $v$ smaller than $m$ on the $i$-th dimension will be stored in a leaf in the left sub-tree of this non-leaf node, while inputs with value $v$ larger than $m$ on the $i$-th dimension will be stored in a leaf in the right sub-tree of this non-leaf node. Each leaf node stores one input value $v$. To be specific, for a set of $k$-dimensional data denoted by $\mathcal{D}$, its KD-Tree is generated as follows: 

\begin{enumerate}
    \item First, denote the root node of the KD-Tree by $R$, find the range of the data in all of the $k$ dimensions, and record the following two values in $R$: the index of the dimension which has the largest range, denoted by $\kappa$, and the median of the data in this dimension, denoted by med. 
    \item Next, divide $\mathcal{D}$ into $\mathcal{D}_l$ and $\mathcal{D}_r$ by the dimension and median recorded in the root, such that all the data points whose value on the dimension $\kappa$ is smaller than or equal to med are in $\mathcal{D}_l$, while the rest are in $\mathcal{D}_r$. 
    \item Denote the left and right child of node $R$ as $R_l$ and $R_r$, respectively. To construct the left (right) sub-tree of $R$, repeat steps 1 \& 2 by regarding $\mathcal{D}_l$ ($\mathcal{D}_r$) as the new $\mathcal{D}$, and $R_l$ ($R_r$) as the new $R$.
    \item Continue with step 3 recursively. When there is only one single data point in the data set $\mathcal{D}$, instead of recording the dimension index and median in the node $R$ as before, the data point itself is stored in $R$, which becomes a leaf node of the KD-Tree.
\end{enumerate}

Herein, to order all the $k$-dimensional locations, we initially create the KD-Tree using these locations. Subsequently, we perform an in-order traversal of the tree. This entails that, upon reaching any node during the traversal, we first record the data from its left subtree, followed by the data of the node itself, and then proceed to capture the data from its right subtree.


\subsection{Other Ordering methods}
Apart from the previous three ordering methods, which are the main focuses of this article, we also implemented some other ordering methods in \textit{ExaGeoStat}, which we describe as follows.

\subsubsection{Maximum-Minimum Degree (MMD) Ordering}
The Maximum-Minimum Degree (MMD) ordering is widely used in solving sparse linear systems. The primary objective of MMD is to rearrange the rows and columns of a given matrix to ensure that each pair of adjacent locations in the final sequence is not excessively close to the original grid layout and to enhance the efficiency of core matrix operations, including matrix-matrix multiplication, factorization, and similar processes. This approach arranges the spatial locations so that each spatial location in the final order is followed by its nearest neighbours. Therefore, in spatial statistics, it is mainly applied in Vecchia approximation (\citealp{guinness}), where we need to find the distribution of each spatial location conditioning on its nearest neighbours. 

The MMD procedure is as follows: Firstly, the algorithm picks the row or column with the maximum number of non-zero entries (highest degree). Then, it identifies the row or column with the minimum degree among the remaining ones. Finally, the algorithm reorders the matrix such that these rows and columns are moved to a position in the matrix, such as the bottom right corner, where their impact on fill-in is minimized.

The covariance matrix integral to MLE operations is dense, contrasting with the sparse matrices the original Maximum-Minimum Degree (MMD) ordering algorithm addresses. To bridge this gap, we tailored the MMD algorithm for implementation in the \textit{ExaGeoStat} software. This adaptation includes defining a user-specified threshold to determine when elements should be treated as non-zero-like or zero-like elements in the original algorithm.

Our preliminary experiments indicate that the modified MMD algorithm is ineffective for the MLE covariance matrix. It results in higher tile ranks than the coordinate-based ordering algorithms like Morton, Hilbert, and KD-Tree. Consequently, due to its limited performance in these initial tests, we have excluded it in the detailed experimental section.

\subsubsection{Graph-Based Ordering Methods}
We also consider adopting some existing ordering methods designed for sparse matrices, which can lower the rank of the matrices. These methods are usually based on the corresponding adjacency graph of the sparse matrix. We cannot directly apply these methods because the covariance matrices are always dense, and in these cases, all the nodes in the adjacency graph will be connected. However, we ``sparsify'' the adjacency graph by setting a threshold on the values in the matrix and letting the edge in the graph be disconnected if its corresponding value in the matrix is smaller than the threshold. In addition, the threshold cannot be too large not to lose too much correlation information. Here, we briefly introduce the two graph-based methods that we conducted experiments with within \textit{ExaGeoStat}: 

\begin{enumerate}
    \item Reversed Cuthill-McKee Algorithm \\~\\
    The Reversed Cuthill-McKee (RCM) Algorithm (\citealp{rcm}) is an algorithm aiming at reducing the bandwidth of a sparse matrix. It works basically in the following steps:
    
\begin{enumerate}[(i)]
    \item Consider the sparse matrix as an adjacency matrix, then form its corresponding graph and start with an arbitrary node in the graph;
    \item Conduct breadth-first search (BFS) traversal on the graph, starting from the arbitrarily chosen node. For each node in the graph, define its level as its distance from the starting node;
    \item Sort all the nodes based on their levels in descending order, while the nodes with the same level are sorted by degree, with nodes having higher degrees appearing first;
    \item Reorder the nodes based on the sorted order and then reverse the order of the reordered nodes to obtain the final ordering.
\end{enumerate}

Although it cannot be guaranteed that the RCM algorithm can lead to the optimal result, experiments have shown that it can significantly reduce the bandwidth of sparse matrices. \\

    \item Minimum Degree Algorithm \\~\\
    Similarly, the Minimum Degree Algorithm (\citealp{george}) also aims at reducing the bandwidth of each row of a sparse matrix, which works in the following steps:
    
\begin{enumerate}[(i)]
    \item Form the graph corresponding to the sparse matrix (considered as an adjacency matrix), then compute the degree of each node in the graph;
    \item Find the node with the minimum degree, then eliminate the selected node by removing it from the graph and updating the degrees of its neighboring nodes;
    \item Update the degrees of the remaining nodes affected by the removal;
    \item Repeat the previous two steps until all nodes have been eliminated.

\end{enumerate}

The order in which the nodes are eliminated forms the new ordering of the vertices, thus yielding the new ordering of the matrix itself.
\end{enumerate}

Our experiments in \textit{ExaGeoStat} showed that these two algorithms could not help accelerate our computation, with many different choices of the thresholds while ``sparsifying'' the covariance matrices. In fact, after reordering the covariance matrices using these two algorithms, the ranks of some of the off-diagonal tiles are still too large, such that the TLR approximation cannot even proceed. 

\section{Spatial Models}
\label{section:model}

\subsection{Univariate Matérn Model}
\label{univmatern}
Most common geostatistical data sets, such as climate and environmental data sets, comprise a collection of locations that are distributed across a specific geographic area, either regularly or irregularly. Each location is linked to a single measurement of a particular climate or environmental variable, such as wind speed, air pressure, soil moisture, or humidity. Often, such data sets are modeled as realizations of Gaussian spatial random fields, as formulated in the introduction section. We denote a realization of a Gaussian random field $Z(\mathbf{s})$ by $\mathbf{Z}=\{Z(\mathbf{s}_1), \dots, Z(\mathbf{s}_n)\}^\top$, where $\mathbf{s}_1, \dots, \mathbf{s}_n$ are spatial locations in $\mathbb{R}^d$ for some $d\in\mathbb{Z}^+$. Without loss of generality, we assume the random field $Z(\mathbf{s})$ has zero mean and a stationary covariance function which can be parametrized by a vector $\boldsymbol{\theta}\in\mathbb{R}^q$ for some $q\geq1$: 
\begin{equation}
    C(\mathbf{h}; \boldsymbol{\theta})=\mbox{Cov}\{Z(\mathbf{s}), Z(\mathbf{s}+\mathbf{h})\},
    \label{staconv}
\end{equation}
where $\mathbf{h}\in\mathbb{R}^d$ is the spatial lag vector, and $C$ is symmetric with respect to $\mathbf{h}$. Herein, the expression (\ref{staconv}) is further simplified from (\ref{covfunc}) due to the stationarity assumption. Therefore, the $(i,j)$-th entry of the covariance matrix $\mathbf{\Sigma}(\boldsymbol{\theta})$ equals $\mathbf{\Sigma}_{ij}(\boldsymbol{\theta})=C(\mathbf{s}_i-\mathbf{s}_j; \boldsymbol{\theta}),\quad i, j=1, \dots, n.$ The log-likelihood function for $\boldsymbol{\theta}$ in this case is formulated in (\ref{loglik}).

In this work, we mainly focus on the Matérn covariance function without nugget effects, and parametrized by $\boldsymbol{\theta}=(\sigma^2, \beta, \nu)^\top$: 
\begin{equation}
\label{maternkernel}
    C_{\mathcal{M}}(d; \boldsymbol{\theta})=\frac{\sigma^2}{\Gamma(\nu)2^{\nu-1}}\left(\frac{d}{\beta}\right)^\nu\mathcal{K}_\nu\left(\frac{d}{\beta}\right),
\end{equation}
where $d=\|\mathbf{s}-\mathbf{s}'\|$ denotes the distance between two locations $\mathbf{s}$ and $\mathbf{s}'$, $\sigma^2, \beta, \nu$ are the variance, range and smoothness parameters, respectively, which all take positive values, and $\mathcal{K}_\nu$ is the modified Bessel function of the second kind of order $\nu$. 

\subsection{Other Spatial Models}

Apart from the most common univariate Matérn model, we may consider some other common spatial models while dealing with spatial data. In \textit{ExaGeoStat}, we also implemented the bivariate Matérn and non-Gaussian models, which we describe as follows.

\subsubsection{Bivariate Matérn Model}
The univariate Matérn covariance function can also be generalized to the multivariate case. Here we introduce the bivariate case (\citealp{gneiting2010matern} \& \citealp{apanasovich2012valid}), which is also implemented in \textit{ExaGeoStat}. The parsimonious bivariate Matérn cross-covariance function between variables $i$ and $j$ is given by \begin{equation}
\label{bivmatern}
    C_{ij}(d; \boldsymbol{\theta})=\frac{\rho_{ij}\sigma_{ii}\sigma_{jj}}{\Gamma(\nu_{ij})2^{\nu_{ij}-1}}\left(\frac{d}{a}\right)^{\nu_{ij}}\mathcal{K}_{\nu_{ij}}\left(\frac{d}{a}\right).
\end{equation}

Similar to \eqref{maternkernel}, here $d=\|\mathbf{s}-\mathbf{s}'\|$ denotes the distance between two locations $\mathbf{s}$ and $\mathbf{s}'$. $i, j=1, 2$ denote the indices of the two components of the bivariate data. The parameter $\boldsymbol{\theta}$ consists of several components. To be specific, $\sigma^2_{11}>0$ and $\sigma^2_{22}>0$ are the marginal variance parameters of the two components; $\alpha>0$ is the spatial range parameter; $\nu_{11}>0$ and $\nu_{22}>0$ are the marginal smoothness parameters of the two components, while $\nu_{12}=\frac{1}{2}(\nu_{11}+\nu_{22})$ is the cross smoothness; $\rho_{ij}$ is the colocated correlation, defined as $$\rho_{ij}=\beta_{ij}\frac{\Gamma(\nu_{ii}+d/2)}{\Gamma(\nu_{ii})}
\frac{\Gamma(\nu_{jj}+d/2)}{\Gamma(\nu_{jj})}\frac{\Gamma(\nu_{ij})}{\Gamma(\nu_{ij}+d/2)},$$ where $\beta_{ii}=\beta_{jj}=1$ and $\beta_{ij}=\beta_{ji}$.

For a bivariate spatial data set with $n$ locations, the size of its corresponding covariance matrix is $2n\times 2n$, with the value of each element given by \eqref{bivmatern}. 

\subsubsection{Non-Gaussian Model}
In many practical studies of spatial data analysis, high skewness, and heavy tails can be captured from the data, which makes it important to go beyond the Gaussian random fields while performing statistical inference. In \textit{ExaGeoStat}, we implemented the Tukey $g$-and-$h$ (TGH) random fields (\citealp{xu2017tukey}), which is a highly flexible non-Gaussian spatial model. The idea of TGH random fields is to distort Gaussian random fields with two parameters, $g$ and $h$, to introduce extra skewness and kurtosis. The TGH random field with location parameter $\xi\in\mathbb{R}$ and scale parameter $\omega>0$ is defined as follows:
\begin{equation*}
    T(\boldsymbol{s})=\xi+\omega\tau_{g, h}\{Z(\boldsymbol{s})\},
\end{equation*}
where $\tau_{g, h}$ is the Tukey's $g$-and-$h$ transformation function $$\tau_{g, h}(z)=g^{-1}\{\exp(gz)-1\}\exp(hz^2/2),$$ and $Z(\boldsymbol{s})$ denotes a standard Gaussian random field which is the same as what we describe in Section \ref{univmatern}. When $g=h=0$, the TGH random fields degenerate to Gaussian random fields. The details of implementing statistical inference for TGH random fields in \textit{ExaGeoStat} can be found in \cite{mondal2022parallel}. 
 
No matter what spatial model we use, our main target is to find the maximum likelihood estimator (MLE) of $\boldsymbol{\theta}$ by maximizing the log-likelihood function shown in (\ref{loglik}). In the next section, we show by numerical experiments how the ordering algorithms applied to the covariance matrices affect this parameter estimation process. Our preliminary experiments show that the effect of the ordering algorithms on spatial data generated from bivariate and non-Gaussian models is similar to those from univariate Matérn models. Therefore, to make it brief, we focus on numerical experiments with spatial data generated from univariate Matérn models in the next section.

\section{Numerical Studies}
In this section, we assess the performance and accuracy of TLR approximation on the spatial covariance matrix $\boldsymbol{\Sigma}(\boldsymbol{\theta})$ using different ordering algorithms. The experiments on small-scale data are performed using an Intel Xeon Gold $6{,}230$ CPU running at $2.10$GHz, with memory size equal to $128$GB, L1, L2 and L3 cache sizes equal to $32$K, $1{,}024$K and $36{,}608$K, respectively; 
while on medium-scale data, we use Intel Xeon E5-2650 v2 CPUs running at $2.60$GHz, with memory size equal to $128$GB, L1, L2 and L3 cache sizes equal to $32$K, $256$K and $20{,}480$K, respectively. 

This section considers spatial data with locations randomly generated within the two-dimensional unit square.

\label{section:exp}
\subsection{Parameter Estimation Accuracy Assessment Using Various Ordering Algorithms}
\label{accuracy}
In this section, we demonstrate the effects of various ordering algorithms on the estimation accuracy of the statistical parameters using a set of synthetic spatial data, which is characterized by a Matérn covariance matrix without nugget effect, as described in equation (\ref{maternkernel}). We first show results from small-scale data ($n=1{,}600$ locations, with tile size $nb=320$) experiments performed on a 40-core shared-memory machine and from medium-scale data ($n=20{,}000$ locations, with tile size $nb=1{,}000$) experiments performed on Shaheen-II supercomputer, using $4$ nodes and $32$ CPU cores on each node. Specifically, we report the parameter estimation results using Morton, Hilbert, and KD-Tree ordering methods and compare them with the accuracy of the estimation if no ordering algorithm is applied. We rely on the same optimization algorithm \texttt{BOBYQA} (\citealp{powell2009bobyqa}), which is a bound-constrained algorithm without using derivatives embedded in \textit{ExaGeoStat}.

\subsubsection{Small-Scale Experiments}
\label{smallscale}
\begin{figure}[htbp]
    \centering
    \includegraphics[width=0.9\linewidth]{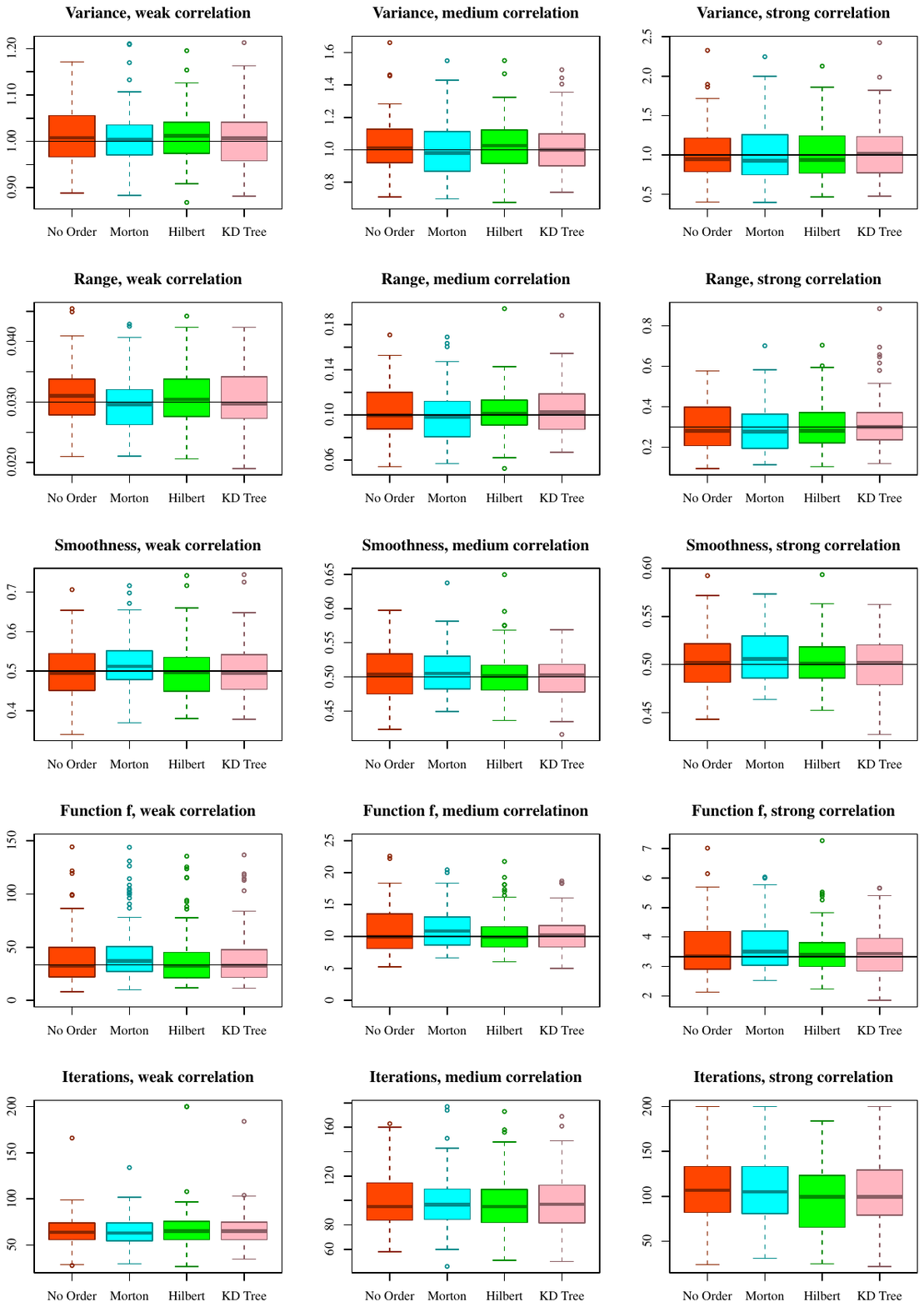}
    \caption{BoxPlots of TLR estimation accuracy with $n=1{,}600$ under either No Order or Morton, Hilbert, KD-Tree orderings. The horizontal lines in the first four rows denote the true values of the corresponding parameters or the function $f$ in (\ref{functionforder}).}
    \label{plot1}
\end{figure}

This section presents a comparative analysis of the effects of three distinct ordering algorithms – Morton, Hilbert, and KD-Tree – on the estimation accuracy of the TLR approximation method, utilizing a covariance matrix of size $1600 \times 1600$. Figure \ref{plot1} shows a set of boxplots representing the estimated values of three parameters (variance, range, and smoothness) for a Matérn covariance function under various correlation structures: weak, medium, and strong. The actual values for these parameters are set at $(\sigma^2=1, \beta=0.03, \nu=0.5)$, $(\sigma^2=1, \beta=0.1, \nu=0.5)$, and $(\sigma^2=1, \beta=0.3, \nu=0.5)$ for each respective structure. The estimation is based on 100 synthetic datasets, each generated using the {\it ExaGeoStat} software. The estimation employed the TLR method with a compression accuracy set to $10^{-7}$ and an optimization tolerance of $10^{-9}$. Additional results under other configurations are available in \ref{append1}. In Figure \ref{plot1}, the initial three rows display the estimated values for variance, range, and smoothness parameters for each setting. The fourth row is dedicated to illustrating the estimated values of the function $f(\sigma^2, \beta, \nu)$ as follows:
\begin{equation}
\label{functionforder}
    f(\sigma^2, \beta, \nu)=\sigma^2\beta^{-2\nu},
\end{equation}
which was proved in \cite{zhang} to be identifiable under infill asymptotics. Moreover, in the fifth row, we plot the number of iterations needed for the optimization process in $\textit{ExaGeoStat}$ to converge to the final results. 

The boxplots show that the Hilbert ordering method yields the most consistent and effective estimation results, demonstrating superior performance in the number of iterations required for convergence. However, the accuracy of parameter estimation using various ordering algorithms is minimally affected by the choice of algorithm. This outcome is advantageous, as our objective is to speed up the optimization process without significantly altering the results.

\subsubsection{Medium-Scale Experiments}
\label{largescale}
\begin{figure}[htbp]
    \centering
    \includegraphics[width=0.9\linewidth]{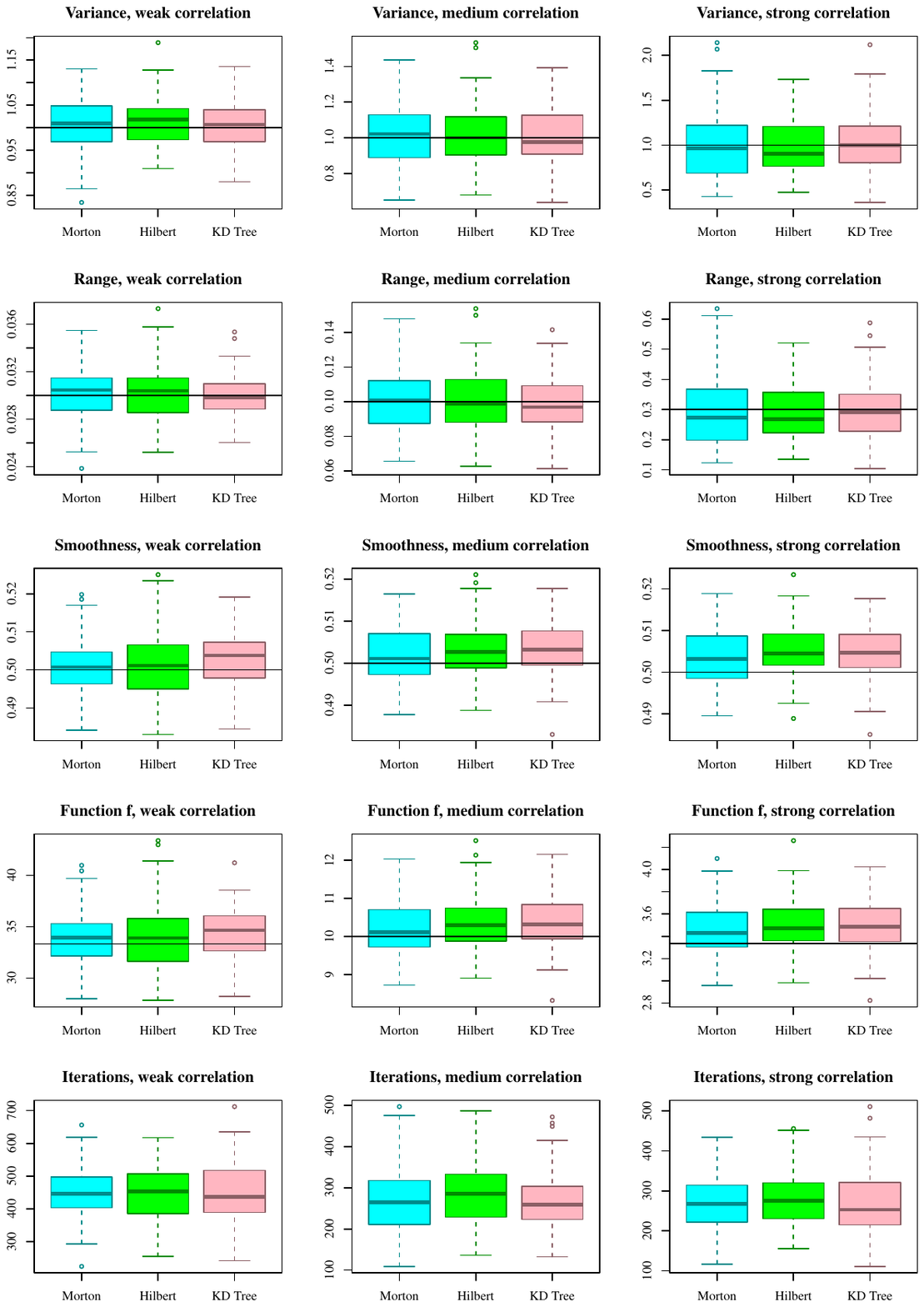}
    \caption{BoxPlots of TLR estimation accuracy with $n=20{,}000$ under Morton, Hilbert, KD-Tree orderings. The horizontal lines in the first four rows denote the true values of the corresponding parameters or the function $f$ in (\ref{functionforder}).}
    \label{plot2}
\end{figure}

In this section, we expand the covariance matrix dimensions to $20,000 \times 20,000$ to examine the impact of different ordering algorithms on a medium-sized correlation matrix. Figure \ref{plot2} compares the outcomes of these algorithms through boxplots. As in the small-scale experiments, we use synthetic spatial data with the Matérn covariance function and weak, medium, and strong correlations under the same settings. Unlike in Figure \ref{plot1}, Figure \ref{plot2} omits the ``no order'' results due to the high ranks of individual tiles, which slows down the estimation process. The experiments are repeated 100 times, each with independently generated synthetic datasets for each setting. The first three rows of Figure \ref{plot2} display the estimation results for the variance, range, and smoothness parameters, while the fourth row depicts the estimated value of the function $f$ as in \eqref{functionforder}. Additionally, the fifth row presents the number of iterations required for the optimization process in \textit{ExaGeoStat} to reach convergence. The boxplots show that the Hilbert ordering method no longer gives superior results in small-scale cases. Morton ordering gives the most stable and unbiased estimation results for most cases. The number of iterations needed to converge is similar among the three algorithms.

\subsection{Tile Ranks with Different Ordering Algorithms}
\label{ranks}
The quality of TLR compression on individual tiles significantly influences the accuracy, memory usage, and computational time required for the TLR approximation algorithm. In this section, we evaluate the effectiveness of TLR compression applied to a covariance matrix generated with a Matérn covariance function. This is achieved by determining the ranks of all matrix tiles while employing various ordering algorithms.

\begin{figure}[htbp]
    \centering
    \includegraphics[width=0.9\linewidth]{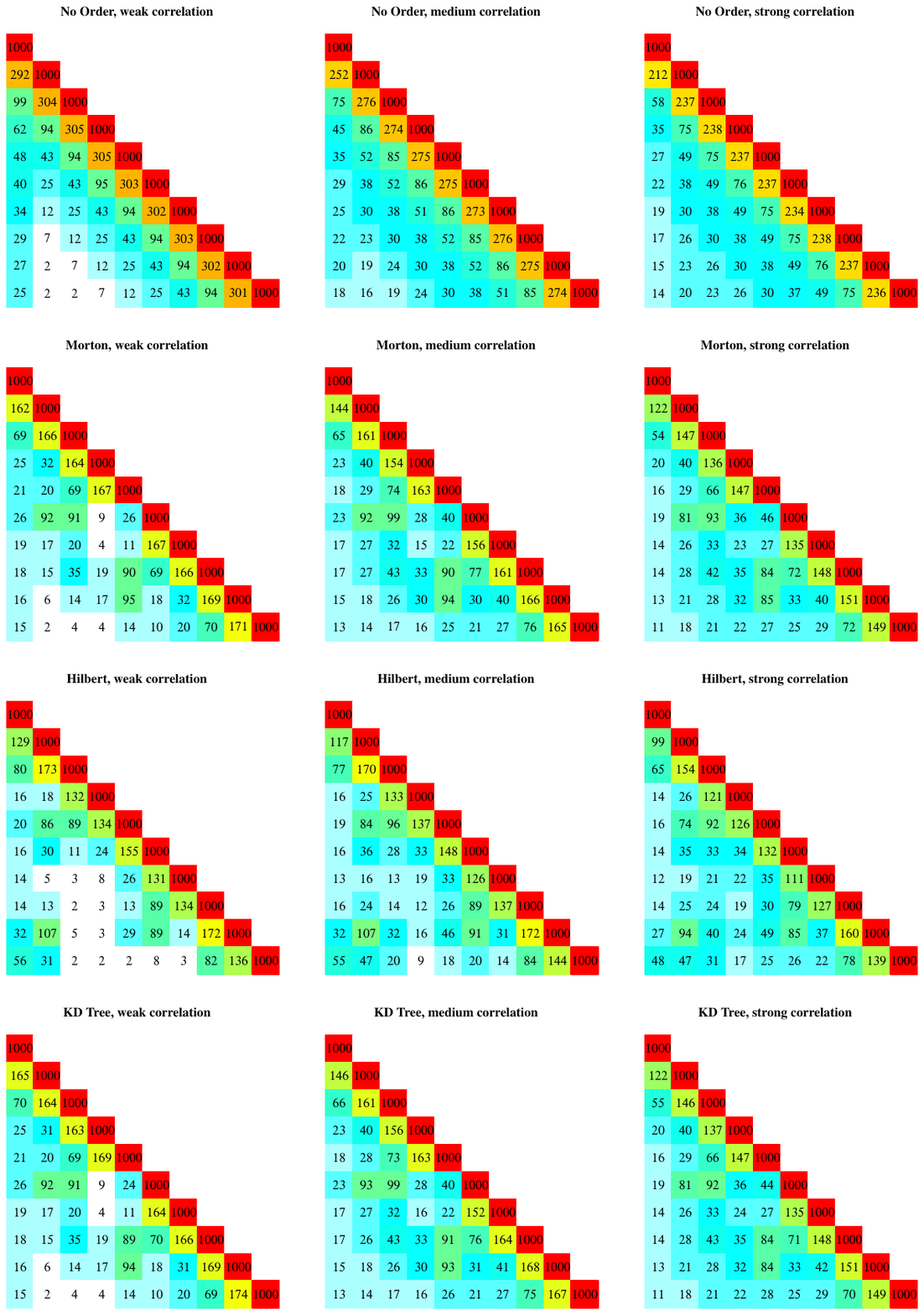}
    \caption{Heatmaps of tile ranks for various ordering algorithms for weak, medium, and strong correlation structures. Each small square symbolizes a $1,000 \times 1,000$ tile, annotated with its corresponding rank. Diagonal tiles maintain full rank. Within each heatmap, darker colors indicate higher ranks.}
    \label{rankplot}
\end{figure}

\begin{figure}[htbp]
    \centering
    \includegraphics[width=0.93\linewidth]{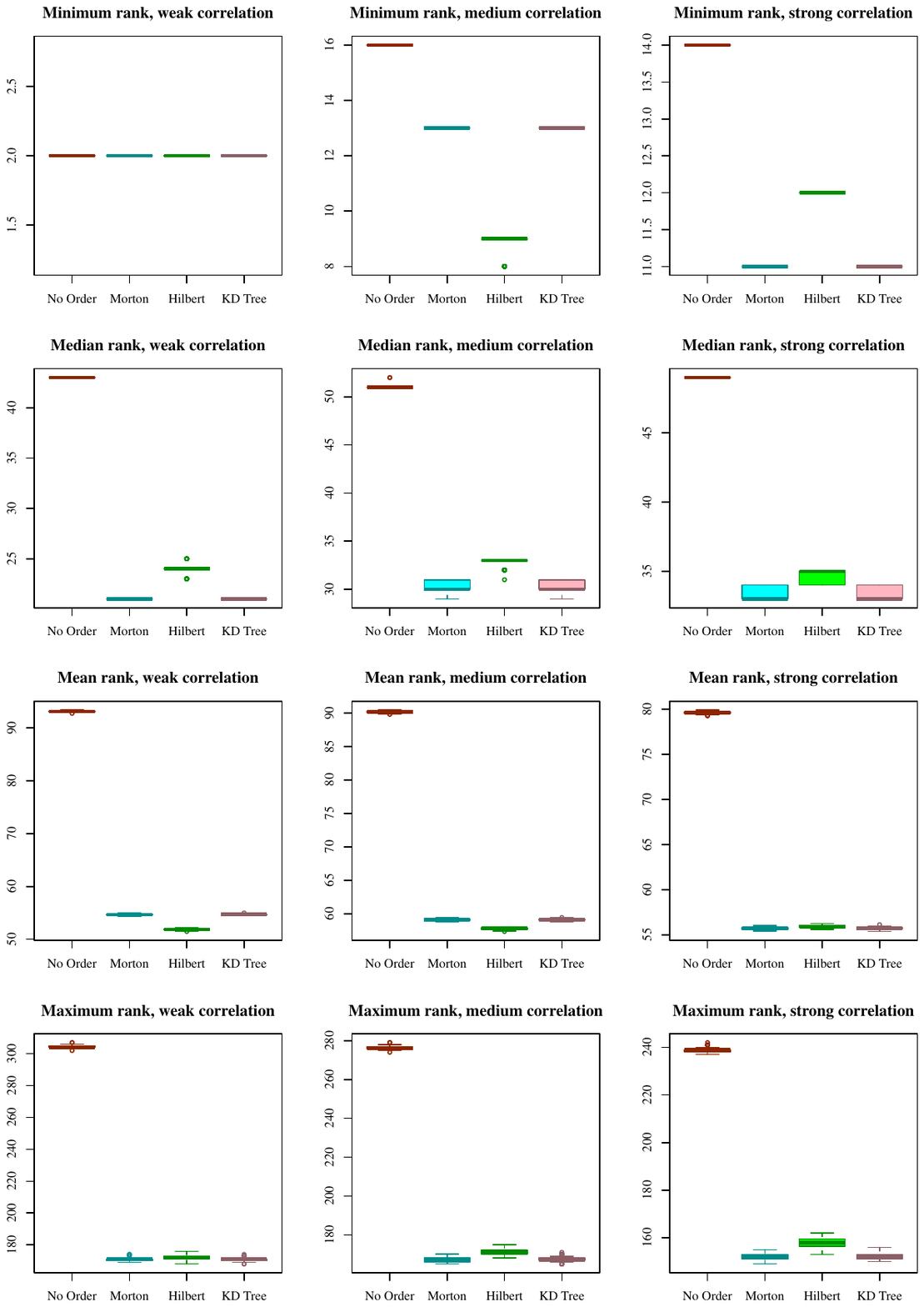}
    \caption{BoxPlots of the minimum, median, mean, and maximum of off-diagonal tile ranks with different ordering methods with weak, medium, and strong correlation structures. We generated $100$ sets of data with $n=10{,}000$ locations, and they are all divided into $1{,}000\times1{,}000$ tiles. }
    \label{rankboxplot}
\end{figure}

\begin{figure}[htbp]
    \centering
    \includegraphics[width=0.93\linewidth]{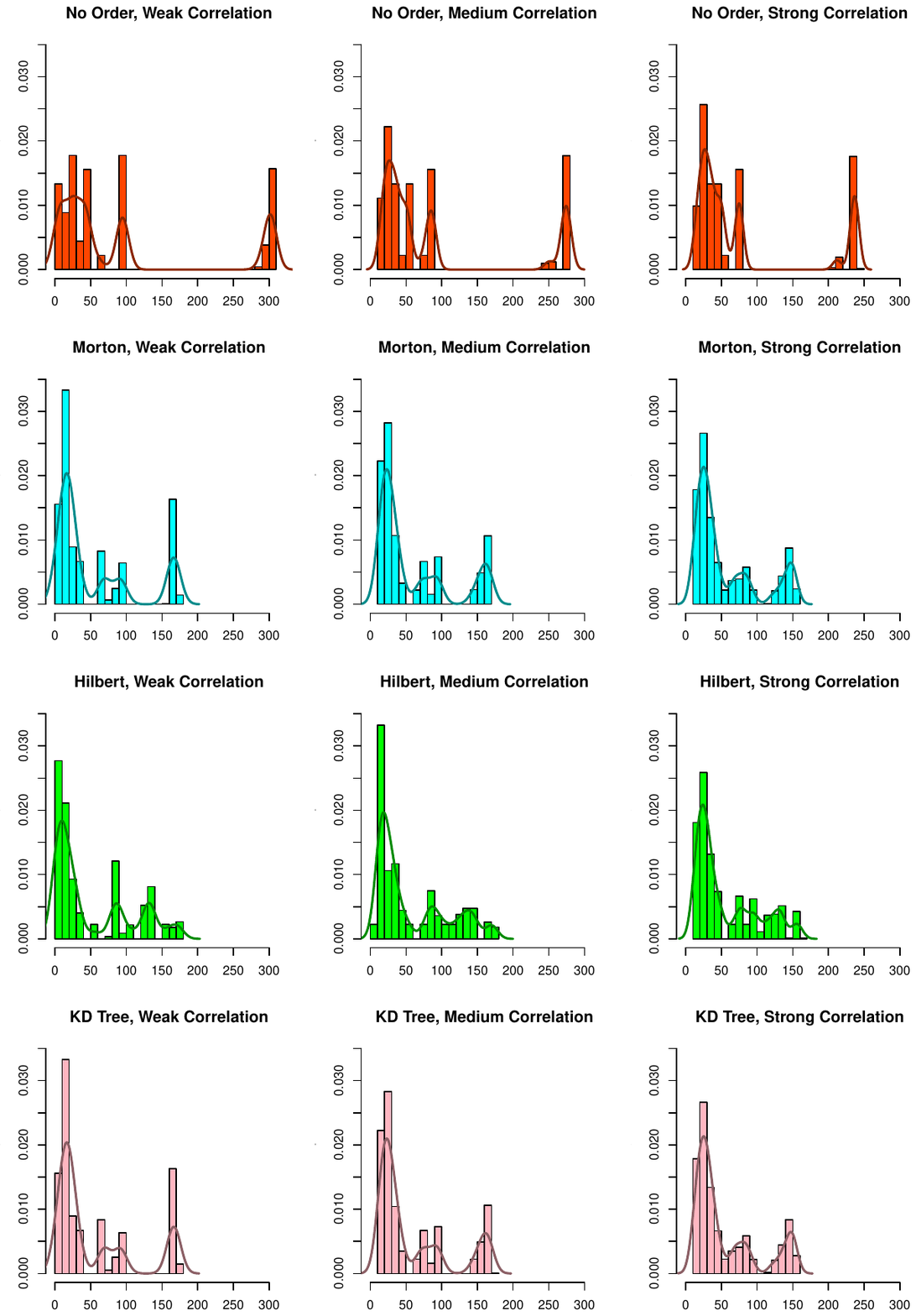}
    \caption{Histograms of the proportions and curves of the empirical densities of off-diagonal tile ranks with different ordering methods with weak, medium, and strong correlation structures. We generated $100$ sets of data with $n=10{,}000$ locations, and they are all divided into $1{,}000\times1{,}000$ tiles.}
    \label{rankhist}
\end{figure}

We rely on synthetic datasets generated with true values $\sigma^2=1, \nu=0.5$ and $\beta=0.03, 0.1, 0.3$, focusing on a location count of $n=10,000$ and a tile size of $nb=1,000$. Additional results under varied settings are shown in \ref{append2}. Figure \ref{rankplot} shows heatmaps of sample covariance matrices for weak, medium, and strong correlation structures. For each correlation type, we created 100 spatial datasets. We also present boxplots of the off-diagonal tiles' minimum, median, mean, and maximum rank values in the corresponding covariance matrices, as shown in Figure \ref{rankboxplot}. Additionally, Figure~\ref{rankhist} presents histograms of these ranks from the 100 datasets. Furthermore, Table \ref{memory} outlines the average memory size needed to store the off-diagonal tiles using different ordering methods for the specified correlation structures. 

\begin{table}[h!]
\centering
\caption{The average memory for storing all the off-diagonal tiles of a covariance matrix of $n=10{,}000$ to $40{,}000$ spatial locations divided into $1{,}000\times1{,}000$ tiles, with different ordering methods and correlation structures. Here, we also compare with the theoretical storage needed without using TLR approximation, denoted by ``dense'', where the storage is the same for different correlation structures if $n$ is fixed, and we can see the storage is much larger than using TLR approximation. The lowest storage required is marked in bold for each number $n$ and each correlation structure. The ``NA'' in this table means that when $n=40{,}000$ with weak correlation, if we do not use any order, then some of the off-diagonal tiles of the covariance matrix will be too dense for the TLR approximation to proceed. \\}
\begin{tabular}{c|l|ccc}
\multicolumn{1}{l|}{}         & Ordering & Weak       & Medium     & Strong     \\ \hline
\multirow{5}{*}{$n=10{,}000$} & Dense    & \multicolumn{3}{c}{$360$MB}          \\
                              & No Order & $67$MB  & $65$MB  & $57$MB  \\
                              & Morton   & $39$MB  & $43$MB  & \textbf{$\textbf{40}$MB}  \\
                              & Hilbert  & \textbf{$\textbf{37}$MB}  & \textbf{$
                              \textbf{42}$MB}  & \textbf{$\textbf{40}$MB}  \\
                              & KD-Tree  & $39$MB  & $43$MB  & \textbf{$\textbf{40}$MB}  \\ \hline
\multirow{5}{*}{$n=20{,}000$} & Dense    & \multicolumn{3}{c}{$1{,}520$MB}      \\
                              & No Order & $256$MB & $250$MB & $227$MB \\
                              & Morton   & $105$MB & $123$MB & $122$MB \\
                              & Hilbert  & \textbf{$\textbf{91}$MB}  & \textbf{$\textbf{110}$MB} & \textbf{$\textbf{113}$MB} \\
                              & KD-Tree  & $100$MB  & $116$MB & $115$MB \\ \hline
\multirow{5}{*}{$n=30{,}000$} & Dense    & \multicolumn{3}{c}{$3{,}480$MB}      \\
                              & No Order & $556$MB & $549$MB & $501$MB \\
                              & Morton   & $191$MB & $233$MB & $237$MB \\
                              & Hilbert  & \textbf{$\textbf{161}$MB} & \textbf{$\textbf{203}$MB} & \textbf{$\textbf{211}$MB} \\
                              & KD-Tree  & $189$MB & $229$MB & $232$MB \\ \hline
\multirow{5}{*}{$n=40{,}000$} & Dense    & \multicolumn{3}{c}{$6{,}240$MB}      \\
                              & No Order & NA         & $955$MB & $870$MB \\
                              & Morton   & $251$MB & $316$MB & \textbf{$\textbf{328}$MB} \\
                              & Hilbert  & \textbf{$\textbf{240}$MB} & \textbf{$\textbf{312}$MB} & $331$MB \\
                              & KD-Tree  & $251$MB & $316$MB & \textbf{$\textbf{328}$MB}
\end{tabular}
\label{memory}
\end{table}

From the figures and the table, we can highlight the performance of the three ordering algorithms (Hilbert, Morton, and KD-Tree) in compressing individual off-diagonal tiles as follows:
\begin{itemize}
    \item {\bf Lower Off-Diagonal Tile Ranks}: All three algorithms successfully reduce off-diagonal tile ranks, which leads to lower memory consumption and higher computation speed.

    \item {\bf Hilbert's Superiority in Weak Correlation}: Hilbert outperforms Morton and KD-Tree in cases with weak correlation. This can be shown by the distribution of off-diagonal tile ranks (Figure \ref{rankhist}), where Morton and KD-Tree show peaks around 150, but Hilbert does not.

    \item {\bf Memory Efficiency}: The memory required to store off-diagonal tiles using Hilbert is smaller than with Morton and KD-Tree in cases of weak or medium correlation (Table \ref{memory}). However, in strong correlation cases, the memory usage is almost identical across all three algorithms.

    \item {\bf Effect of Correlation Strength on Tile Ranks}: Figure \ref{rankplot} shows that the ranks of off-diagonal tiles become higher when the correlation becomes weaker. At first, this may seem counter-intuitive, but the explanation is that the covariance matrix values change more rapidly towards the off-diagonal direction when the correlation is smaller, which results in higher tile ranks.
\end{itemize}

\subsection{Computation Performance Assessment}
This section assesses the computation performance of Morton, Hilbert, and KD-Tree ordering algorithms under different correlation scenarios. 

\subsubsection{Cholesky Factorization Performance}
Recalling the formula of the log-likelihood in \eqref{loglik}, the Cholesky factorization of the covariance matrix $\boldsymbol \Sigma$ is the most time-consuming operation while calculating the log-likelihood in each iteration during the optimization process to find the MLE. Hence, this process is a key indicator of the overall performance.

\begin{figure}[h!]
    \centering
    \includegraphics[width=\linewidth]{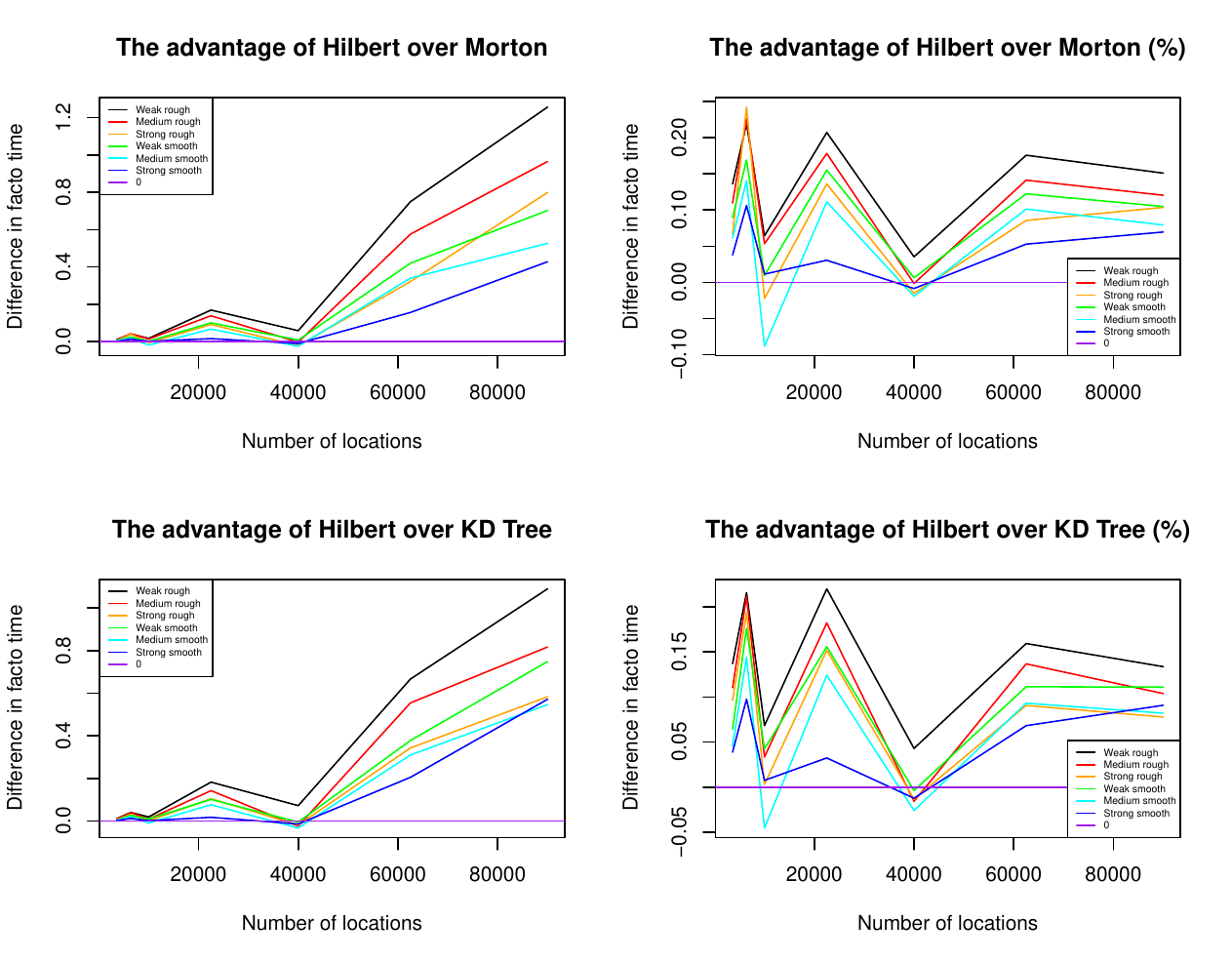}
    \caption{TLR Cholesky Factorization Execution Time. Variance set at $\sigma^2=1$. The weak, medium, and strong refer to the range parameter $\beta$ values of $0.03, 0.1$, and $0.3$, respectively. The words ``rough'' and ``smooth'' correspond to the cases where $\nu=0.5$ and $\nu=1$, respectively. The ``facto time'' on the y-axis indicates the time taken for Cholesky factorization. The left-hand plots show the actual time differences (in seconds) for various orderings, while the right-hand plots present these differences in percentage terms.}
    \label{perf1}
\end{figure}

In Figure \ref{perf1}, we compare the execution time of a single TLR Cholesky factorization of a compressed covariance matrix using the three ordering algorithms under different settings. The corresponding TLR accuracy is $10^{-7}$, and the tile size is $1000$. We consider the cases where the number of locations $n$ equals to $3{,}600$, $6{,}400$, $10{,}000$, $22{,}500$, $40{,}000$, $62{,}500$ and $90{,}000$, and the corresponding results are shown in each subfigure of Figure \ref{perf1} from left to right in the $x$-axis. 

The results indicate that TLR Cholesky factorization using Hilbert ordering outperforms Morton or KD-Tree ordering in most cases. Notably, the efficiency of Hilbert ordering compared to the other two increases with the scaling up of the data. We also note that the difference in the factorization time using different ordering methods is more significant when the correlation among the spatial data is weaker, which is in line with our finding in Section \ref{ranks} that when the correlation is weaker, Hilbert ordering can reduce the ranks of off-diagonal tiles as well as the memory size needed to store the covariance matrices. In addition, when the smoothness $\nu=0.5$, we observe a larger difference between Hilbert's performance and other orderings, compared with the case where $\nu=1$.

\subsection{Application on Soil Moisture Data}
\label{realdata}


We apply our proposed methods on a daily soil moisture data set, which was organized by \cite{litvinenko2019likelihood}, and can be downloaded from \url{https://github.com/litvinen/HLIBCov.git}. Soil moisture plays a crucial role in assessing the condition of hydrological processes and finds widespread applications in weather forecasting, crop yield prediction, and early detection of flood and drought events. Improving the characterization of soil moisture has demonstrated a significant enhancement in weather forecasting. However, the high spatial resolution needed often results in large datasets from numerical models, making the computation of many statistical inference methods impractical. In this experiment, we examine high-resolution soil moisture data collected from January 1, 2014, in the upper soil layer of the Mississippi River basin in the United States. The spatial resolution of the data is $0.0083$ degrees, and the distance of one-degree difference in this region is approximately $87.5$ km. The grid encompasses $2{,}432{,}070$ locations, with $2{,}000{,}000$ observations and $432{,}070$ missing values. As a result, the spatial data available do not conform to a regular grid. In our experiment, subsets of the data without missing values with numbers of locations $n=2{,}000, 4{,}000, 8{,}000, 16{,}000, 32{,}000, 64{,}000$ are randomly selected. We apply Morton, Hilbert, and KD-Tree orderings with TLR approximation to perform estimation on the variance parameter $\sigma^2$, range $\beta$ and smoothness $\nu$ assuming a Gaussian random field with Matérn covariance function as defined in (\ref{maternkernel}).

\begin{table}[h!]
\centering
\caption{Experimental results on soil moisture data for various numbers of locations, $n$. Parameter estimation results are shown in the columns with titles $\hat \sigma^2$, $\hat \beta$, and $\hat \nu$, respectively, and in the last three columns, we show the number of iterations needed until the optimization process converges, as well as the average execution time for each iteration and the total execution time of the optimization, both in seconds. We do not present the results using any order because, in that case, some off-diagonal tiles will be too dense for the TLR approximation to proceed. \\}
\begin{tabular}{l|l|ccc|ccc}
\multicolumn{1}{c|}{\multirow{2}{*}{$n$}}        & \multicolumn{1}{c|}{\multirow{2}{*}{Ordering}} & \multicolumn{1}{c}{\multirow{2}{*}{$\hat \sigma^2$}} & \multirow{2}{*}{$\hat \beta$} & \multirow{2}{*}{$\hat \nu$} & Number of            & Time to              & Time per             \\
\multicolumn{1}{c|}{}                            & \multicolumn{1}{c|}{}                          & \multicolumn{1}{c}{}                            &                          &                        & Iterations           & Solution          & Iteration        \\ \hline
\multirow{3}{*}{$2{,}000$}  & Morton   & $1.1541$   & $0.2335$ & $0.2655$ & $416$  & $80$s                                                        & $0.19$s                                                                              \\
                            & Hilbert  & $1.1549$   & $0.2339$ & $0.2655$ & $383$  & $70$s                                                        & $0.18$s                                                                              \\
                            & KD-Tree  & $1.1552$   & $0.2340$ & $0.2655$ & $435$  & $83$s                                                        & $0.19$s                                                                              \\ \cline{1-8}
\multirow{3}{*}{$4{,}000$}  & Morton   & $1.0602$   & $0.2640$ & $0.2350$ & $397$  & $135$s                                                       & $0.34$s                                                                              \\
                            & Hilbert  & $1.0602$   & $0.2640$ & $0.2350$ & $355$  & $118$s                                                       & $0.33$s                                                                             \\
                            & KD-Tree  & $1.0600$   & $0.2639$ & $0.2350$ & $310$  & $103$s                                                       & $0.33$s                                                                              \\ \cline{1-8}
\multirow{3}{*}{$8{,}000$}  & Morton   & $1.0637$   & $0.2318$ & $0.2390$ & $566$  & $1{,}807$s                                                      & $3.19$s                                                                              \\
                            & Hilbert  & $1.0638$   & $0.2318$ & $0.2390$ & $531$  & $1{,}707$s                                                      & $3.21$s                                                                               \\
                            & KD-Tree  & $1.0636$   & $0.2318$ & $0.2389$ & $493$  & $1{,}558$s                                                      & $3.16$s                                                                              \\ \cline{1-8}
\multirow{3}{*}{$16{,}000$} & Morton   & $1.0725$   & $0.2517$ & $0.2353$ & $203$  & $2{,}242$s                                                      & $11.04$s                                                                              \\
                            & Hilbert  & $1.0678$   & $0.2493$ & $0.2353$ & $282$  & $3{,}085$s                                                      & $10.94$s                                                                              \\
                            & KD-Tree  & $1.0680$   & $0.2492$ & $0.2354$ & $293$  & $3{,}175$s                                                      & $10.84$s                                                                              \\ \cline{1-8}
\multirow{3}{*}{$32{,}000$} & Morton   & $1.0245$   & $0.1577$ & $0.2610$ & $764$  & $20{,}524$s                                                     & $26.86$s                                                                              \\
                            & Hilbert  & $1.0261$   & $0.1582$ & $0.2610$ & $605$ & $15{,}550$s  & $25.70$s                                     \\
                              & KD-Tree  & $1.0257$   & $0.1581$ & $0.2610$ & $519$   & $13{,}101$s & $25.24$s
\end{tabular}
\label{realres}
\end{table}

In Table \ref{realres}, we show the estimation results for the parameters $\sigma^2$, $\beta$, and $\nu$, as well as the number of iterations needed until the optimization process converges, the average execution time for each iteration, and the total execution time of the optimization, which are shown in the last three columns. Different ordering methods have almost no influence on the parameter estimation results. For the time it takes to complete each iteration, there is also no big difference among these three ordering algorithms. This makes sense since, from the parameter estimation results, we can see the dependence structure of this data set is between medium and strong, closer to the strong side. As we demonstrated in Table \ref{memory} and Figure \ref{rankhist}, the difference in reduction of off-diagonal tiles among different ordering algorithms is not that much in this case. 

\section{Conclusion and Discussion}
\label{section:con}
In this work, we implemented several ordering algorithms in the \textit{ExaGeoStat} framework to re-order the locations of spatial data before generating the spatial covariance matrices and performing Tile Low-Rank (TLR) approximation. Some ordering methods, namely Morton, Hilbert, and KD-Tree, can largely reduce computation time and storage while performing parameter estimation on the data using Maximum Likelihood Estimation. We conducted numerical experiments with data generated from Gaussian random fields with the Mat\'ern covariance function. To be specific, we examined the accuracy of parameter estimation and the convergence rate (Figures \ref{plot1}, \ref{plot2}), the ranks of off-diagonal tiles in TLR estimation (Figures \ref{rankplot}, \ref{rankboxplot}, \ref{rankhist}) as well as the storage required (Table \ref{memory}), and the execution time of Cholesky factorization of the covariance matrix (Figure \ref{perf1}). In addition, we applied our methods to a soil moisture data set (Table \ref{realres}).

From our conducted numerical experiments, the ordering algorithms do not significantly affect the parameter estimation accuracy. However, slight differences can be observed: in small-scale cases, Hilbert outperforms the other ordering methods, while in medium-scale cases, Morton achieves the highest accuracy. The computation time and storage of the estimation procedure varies depending on the ordering algorithms employed, especially when the correlation structure is weak; Hilbert can reduce the ranks of off-diagonal tiles to the largest extent and, therefore, reduce the computation time. When the correlation among the spatial data gets stronger and stronger, Hilbert's advantage gradually vanishes. This observation was also verified by our experiments on real data. 

\section*{Acknowledgements}
The authors thank the Extreme Computing Research Center (ECRC) at King Abdullah University of Science and Technology (KAUST). This research utilized CPU-based systems and the Shaheen II supercomputer, housed at the Supercomputing Laboratory at KAUST, for computational resources.


\bibliographystyle{asa}
\bibliography{bibliography}

\begin{thebibliography}{37}
\newcommand{\enquote}[1]{``#1''}
\expandafter\ifx\csname natexlab\endcsname\relax\def\natexlab#1{#1}\fi

\bibitem[{Abdulah et~al.(2018{\natexlab{a}})Abdulah, Ltaief, Sun, Genton, and Keyes}]{exageostat}
Abdulah, S., Ltaief, H., Sun, Y., Genton, M.~G., and Keyes, D.~E. (2018{\natexlab{a}}), \enquote{Exageostat: A high performance unified software for geostatistics on manycore systems,} \textit{IEEE Transactions on Parallel and Distributed Systems}, 29, 2771--2784.

\bibitem[{Abdulah et~al.(2018{\natexlab{b}})Abdulah, Ltaief, Sun, Genton, and Keyes}]{abdulah2018parallel}
--- (2018{\natexlab{b}}), \enquote{Parallel approximation of the maximum likelihood estimation for the prediction of large-scale geostatistics simulations,} in \textit{2018 IEEE international conference on cluster computing (CLUSTER)}, IEEE, pp. 98--108.

\bibitem[{Akbudak et~al.(2017)Akbudak, Ltaief, Mikhalev, and Keyes}]{akbudak2017tile}
Akbudak, K., Ltaief, H., Mikhalev, A., and Keyes, D. (2017), \enquote{Tile low rank Cholesky factorization for climate/weather modeling applications on manycore architectures,} in \textit{International Conference on High Performance Computing}, Springer, pp. 22--40.

\bibitem[{Apanasovich et~al.(2012)Apanasovich, Genton, and Sun}]{apanasovich2012valid}
Apanasovich, T.~V., Genton, M.~G., and Sun, Y. (2012), \enquote{A valid Mat{\'e}rn class of cross-covariance functions for multivariate random fields with any number of components,} \textit{Journal of the American Statistical Association}, 107, 180--193.

\bibitem[{Augonnet et~al.(2009)Augonnet, Thibault, Namyst, and Wacrenier}]{augonnet2009starpu}
Augonnet, C., Thibault, S., Namyst, R., and Wacrenier, P.-A. (2009), \enquote{StarPU: a unified platform for task scheduling on heterogeneous multicore architectures,} in \textit{Euro-Par 2009 Parallel Processing: 15th International Euro-Par Conference, Delft, The Netherlands, August 25-28, 2009. Proceedings 15}, Springer, pp. 863--874.

\bibitem[{Banerjee et~al.(2008)Banerjee, Gelfand, Finley, and Sang}]{banerjee2008gaussian}
Banerjee, S., Gelfand, A.~E., Finley, A.~O., and Sang, H. (2008), \enquote{Gaussian predictive process models for large spatial data sets,} \textit{Journal of the Royal Statistical Society Series B: Statistical Methodology}, 70, 825--848.

\bibitem[{Bentley(1975)}]{kdtree}
Bentley, J.~L. (1975), \enquote{Multidimensional binary search trees used for associative searching,} \textit{Communications of the ACM}, 18, 509--517.

\bibitem[{Bosilca et~al.(2013)Bosilca, Bouteiller, Danalis, Faverge, H{\'e}rault, and Dongarra}]{bosilca2013parsec}
Bosilca, G., Bouteiller, A., Danalis, A., Faverge, M., H{\'e}rault, T., and Dongarra, J.~J. (2013), \enquote{Parsec: Exploiting heterogeneity to enhance scalability,} \textit{Computing in Science \& Engineering}, 15, 36--45.

\bibitem[{Cao et~al.(2022)Cao, Abdulah, Alomairy, Pei, Nag, Bosilca, Dongarra, Genton, Keyes, Ltaief, et~al.}]{cao2022reshaping}
Cao, Q., Abdulah, S., Alomairy, R., Pei, Y., Nag, P., Bosilca, G., Dongarra, J., Genton, M.~G., Keyes, D.~E., Ltaief, H., et~al. (2022), \enquote{Reshaping geostatistical modeling and prediction for extreme-scale environmental applications,} in \textit{SC22: International Conference for High Performance Computing, Networking, Storage and Analysis}, IEEE, pp. 1--12.

\bibitem[{Chen et~al.(2021)Chen, Genton, and Sun}]{chen2021space}
Chen, W., Genton, M.~G., and Sun, Y. (2021), \enquote{Space-time covariance structures and models,} \textit{Annual Review of Statistics and Its Application}, 8, 191--215.

\bibitem[{Cressie and Johannesson(2008)}]{cressie2008fixed}
Cressie, N. and Johannesson, G. (2008), \enquote{Fixed rank kriging for very large spatial data sets,} \textit{Journal of the Royal Statistical Society Series B: Statistical Methodology}, 70, 209--226.

\bibitem[{Cuthill and McKee(1969)}]{rcm}
Cuthill, E. and McKee, J. (1969), \enquote{Reducing the bandwidth of sparse symmetric matrices,} in \textit{Proceedings of the 1969 24th national conference}, pp. 157--172.

\bibitem[{Duran et~al.(2011)Duran, Ayguad{\'e}, Badia, Labarta, Martinell, Martorell, and Planas}]{duran2011ompss}
Duran, A., Ayguad{\'e}, E., Badia, R.~M., Labarta, J., Martinell, L., Martorell, X., and Planas, J. (2011), \enquote{Ompss: a proposal for programming heterogeneous multi-core architectures,} \textit{Parallel Processing Letters}, 21, 173--193.

\bibitem[{Edwards et~al.(2014)Edwards, Trott, and Sunderland}]{edwards2014kokkos}
Edwards, H.~C., Trott, C.~R., and Sunderland, D. (2014), \enquote{Kokkos: Enabling manycore performance portability through polymorphic memory access patterns,} \textit{Journal of Parallel and Distributed Computing}, 74, 3202--3216.

\bibitem[{Elhorst et~al.(2021)Elhorst, Gross, and Tereanu}]{elhorst2021cross}
Elhorst, J.~P., Gross, M., and Tereanu, E. (2021), \enquote{Cross-sectional dependence and spillovers in space and time: Where spatial econometrics and global VAR models meet,} \textit{Journal of Economic Surveys}, 35, 192--226.

\bibitem[{George and Liu(1989)}]{george}
George, A. and Liu, J.~W. (1989), \enquote{The evolution of the minimum degree ordering algorithm,} \textit{SIAM Review}, 31, 1--19.

\bibitem[{Gneiting et~al.(2007)Gneiting, Genton, and Guttorp}]{gneiting2006geostatistical}
Gneiting, T., Genton, M.~G., and Guttorp, P. (2007), \enquote{Geostatistical space-time models, stationarity, separability, and full symmetry,} \textit{Monographs On Statistics and Applied Probability}, 107, 151--175.

\bibitem[{Gneiting et~al.(2010)Gneiting, Kleiber, and Schlather}]{gneiting2010matern}
Gneiting, T., Kleiber, W., and Schlather, M. (2010), \enquote{Mat{\'e}rn cross-covariance functions for multivariate random fields,} \textit{Journal of the American Statistical Association}, 105, 1167--1177.

\bibitem[{Guinness(2018)}]{guinness}
Guinness, J. (2018), \enquote{Permutation and grouping methods for sharpening Gaussian process approximations,} \textit{Technometrics}, 60, 415--429.

\bibitem[{Hilbert(1935)}]{hilbert}
Hilbert, D. (1935), \enquote{{\"U}ber die stetige Abbildung einer Linie auf ein Fl{\"a}chenst{\"u}ck,} in \textit{Dritter Band: Analysis{\textperiodcentered} Grundlagen der Mathematik{\textperiodcentered} Physik Verschiedenes}, Springer, pp. 1--2.

\bibitem[{Kale and Krishnan(1993)}]{kale1993charm++}
Kale, L.~V. and Krishnan, S. (1993), \enquote{Charm++ a portable concurrent object oriented system based on c++,} in \textit{Proceedings of the eighth annual conference on Object-oriented programming systems, languages, and applications}, pp. 91--108.

\bibitem[{Kaufman et~al.(2008)Kaufman, Schervish, and Nychka}]{kaufman2008covariance}
Kaufman, C.~G., Schervish, M.~J., and Nychka, D.~W. (2008), \enquote{Covariance tapering for likelihood-based estimation in large spatial data sets,} \textit{Journal of the American Statistical Association}, 103, 1545--1555.

\bibitem[{Kim and Gu(2004)}]{kim2004smoothing}
Kim, Y.-J. and Gu, C. (2004), \enquote{Smoothing spline Gaussian regression: more scalable computation via efficient approximation,} \textit{Journal of The Royal Statistical Society Series B: Statistical Methodology}, 66, 337--356.

\bibitem[{Litvinenko et~al.(2019)Litvinenko, Sun, Genton, and Keyes}]{litvinenko2019likelihood}
Litvinenko, A., Sun, Y., Genton, M.~G., and Keyes, D.~E. (2019), \enquote{Likelihood approximation with hierarchical matrices for large spatial datasets,} \textit{Computational Statistics \& Data Analysis}, 137, 115--132.

\bibitem[{Mondal et~al.(2022)Mondal, Abdulah, Ltaief, Sun, Genton, and Keyes}]{mondal2022parallel}
Mondal, S., Abdulah, S., Ltaief, H., Sun, Y., Genton, M.~G., and Keyes, D.~E. (2022), \enquote{Parallel approximations of the Tukey g-and-h likelihoods and predictions for non-Gaussian geostatistics,} in \textit{2022 IEEE International Parallel and Distributed Processing Symposium (IPDPS)}, IEEE, pp. 379--389.

\bibitem[{Moraga and Montes(2011)}]{moraga2011detection}
Moraga, P. and Montes, F. (2011), \enquote{Detection of spatial disease clusters with LISA functions,} \textit{Statistics in Medicine}, 30, 1057--1071.

\bibitem[{Morton(1966)}]{morton}
Morton, G. (1966), \textit{A computer oriented geodetic data base and a new technique in file sequencing}, IBM Ltd., Ottawa.

\bibitem[{Ombao et~al.(2008)Ombao, Shao, Rykhlevskaia, Fabiani, and Gratton}]{ombao2008spatio}
Ombao, H., Shao, X., Rykhlevskaia, E., Fabiani, M., and Gratton, G. (2008), \enquote{Spatio-spectral analysis of brain signals,} \textit{Statistica Sinica}, 1465--1482.

\bibitem[{Powell et~al.(2009)}]{powell2009bobyqa}
Powell, M.~J. et~al. (2009), \enquote{The BOBYQA algorithm for bound constrained optimization without derivatives,} \textit{Cambridge NA Report NA2009/06, University of Cambridge, Cambridge}, 26.

\bibitem[{Rue and Held(2005)}]{rue2005gaussian}
Rue, H. and Held, L. (2005), \textit{Gaussian Markov Random Fields: Theory and Applications}, CRC press.

\bibitem[{Rue and Tjelmeland(2002)}]{rue2002fitting}
Rue, H. and Tjelmeland, H. (2002), \enquote{Fitting Gaussian Markov random fields to Gaussian fields,} \textit{Scandinavian Journal of Statistics}, 29, 31--49.

\bibitem[{Sinopoli et~al.(2004)Sinopoli, Schenato, Franceschetti, Poolla, Jordan, and Sastry}]{sinopoli2004kalman}
Sinopoli, B., Schenato, L., Franceschetti, M., Poolla, K., Jordan, M.~I., and Sastry, S.~S. (2004), \enquote{Kalman filtering with intermittent observations,} \textit{IEEE Transactions on Automatic Control}, 49, 1453--1464.

\bibitem[{Skilling(2004)}]{skilling}
Skilling, J. (2004), \enquote{Programming the Hilbert curve,} in \textit{AIP Conference Proceedings}, American Institute of Physics, vol. 707, pp. 381--387.

\bibitem[{Sun et~al.(2015)Sun, Bowman, Genton, and Tokay}]{sun2015matern}
Sun, Y., Bowman, K.~P., Genton, M.~G., and Tokay, A. (2015), \enquote{A Mat{\'e}rn model of the spatial covariance structure of point rain rates,} \textit{Stochastic Environmental Research and Risk Assessment}, 29, 411--416.

\bibitem[{Sun et~al.(2012)Sun, Li, and Genton}]{sun2012geostatistics}
Sun, Y., Li, B., and Genton, M.~G. (2012), \enquote{Geostatistics for large datasets,} in \textit{Advances and Challenges in Space-time Modelling of Natural Events}, Springer, pp. 55--77.

\bibitem[{Xu and Genton(2017)}]{xu2017tukey}
Xu, G. and Genton, M.~G. (2017), \enquote{Tukey g-and-h random fields,} \textit{Journal of the American Statistical Association}, 112, 1236--1249.

\bibitem[{Zhang(2004)}]{zhang}
Zhang, H. (2004), \enquote{Inconsistent estimation and asymptotically equal interpolations in model-based geostatistics,} \textit{Journal of the American Statistical Association}, 99, 250--261.

\end{thebibliography}


\appendix
\section{Supplementary Experimental Results}
In Sections \ref{accuracy} and \ref{ranks}, we analyzed the parameter estimation accuracy and the effect of TLR compression to a given covariance matrix generated using a Matérn covariance kernel, with synthetic data sets generated using rough settings (i.e., the smoothness parameter $\nu=0.5$). In the appendix, we show some results using smooth settings ($\nu=1$), and in this case, for the weak, medium and strong correlation structures, the true values of the variance, range and smoothness parameters are set to be $(\sigma^2=1, \beta=0.025, \nu=1)$, $(\sigma^2=1, \beta=0.075, \nu=1)$, and $(\sigma^2=1, \beta=0.2, \nu=1)$, respectively. 

\subsection{Parameter Estimation}
\label{append1}

\begin{figure}[htbp]
    \centering
    \includegraphics[width=0.9\linewidth]{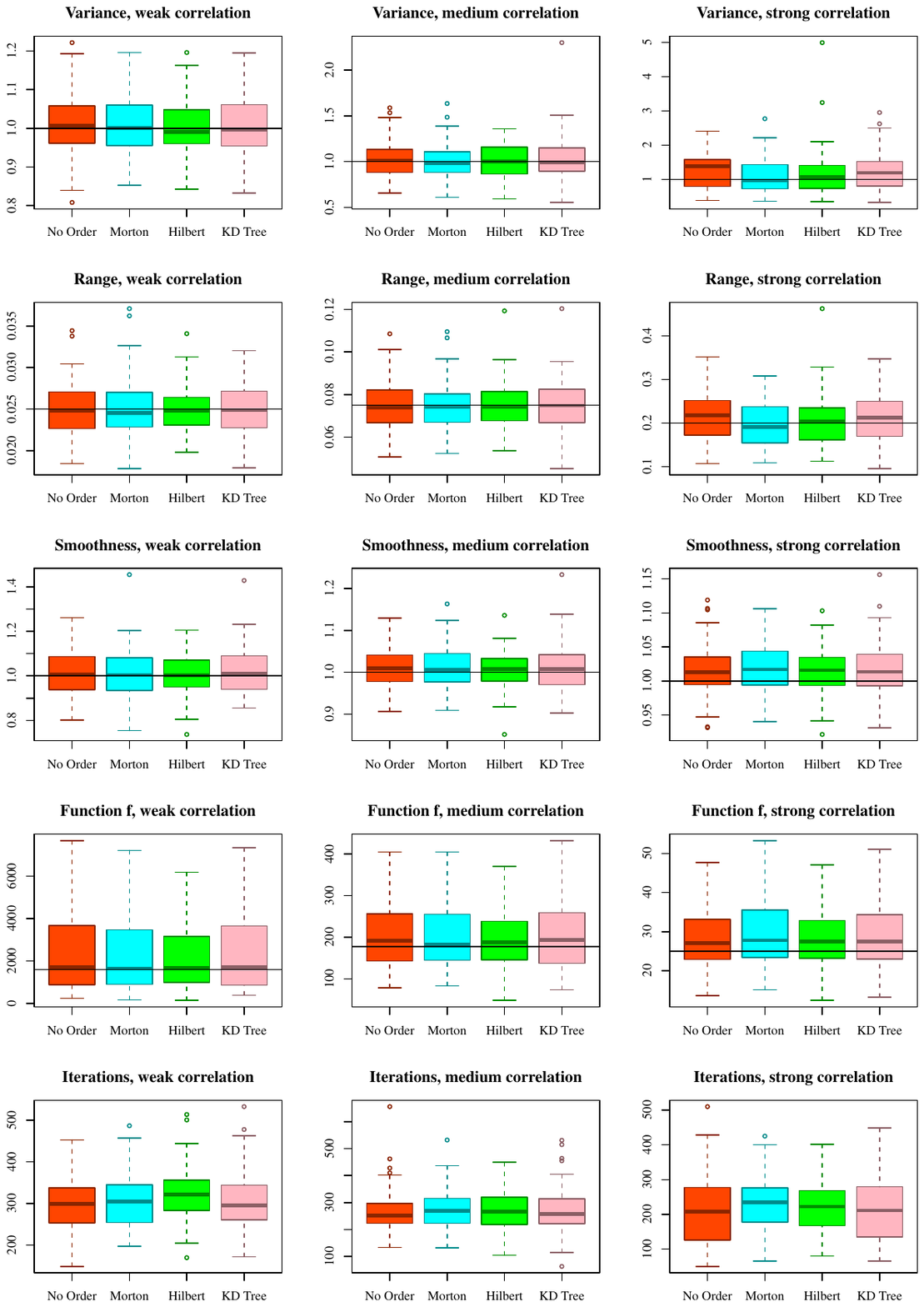}
    \caption{BoxPlots of TLR estimation accuracy with $n=1{,}600$ under either No Order, or Morton, Hilbert, KD-Tree orderings. The data are generated from smooth settings. The horizontal lines in the first four rows denote the true values of the corresponding parameters or the function $f$ in (\ref{functionforder}).}
    \label{smooth}
\end{figure}

In Figure \ref{smooth}, we demonstrate the parameter estimation results of small-scale data ($n=1{,}600$ locations, with tile size $nb=320$) experiments performed on a normal machine with $40$ CPU core, with data generated from the smooth settings. As in Section \ref{smallscale}, the experiments are repeated $100$ times using different synthetic data sets generated independently each time for each setting. In the first three rows, we show the estimation results of the variance, range, and smoothness parameters in each setting, while in the fourth row, we plot the estimated value of the function $f$ defined in (\ref{functionforder}), to give an overall evaluation of the accuracy of the three parameters. 

\subsection{Tile Ranks}
\label{append2}

\begin{figure}[htbp]
    \centering
    \includegraphics[width=0.9\linewidth]{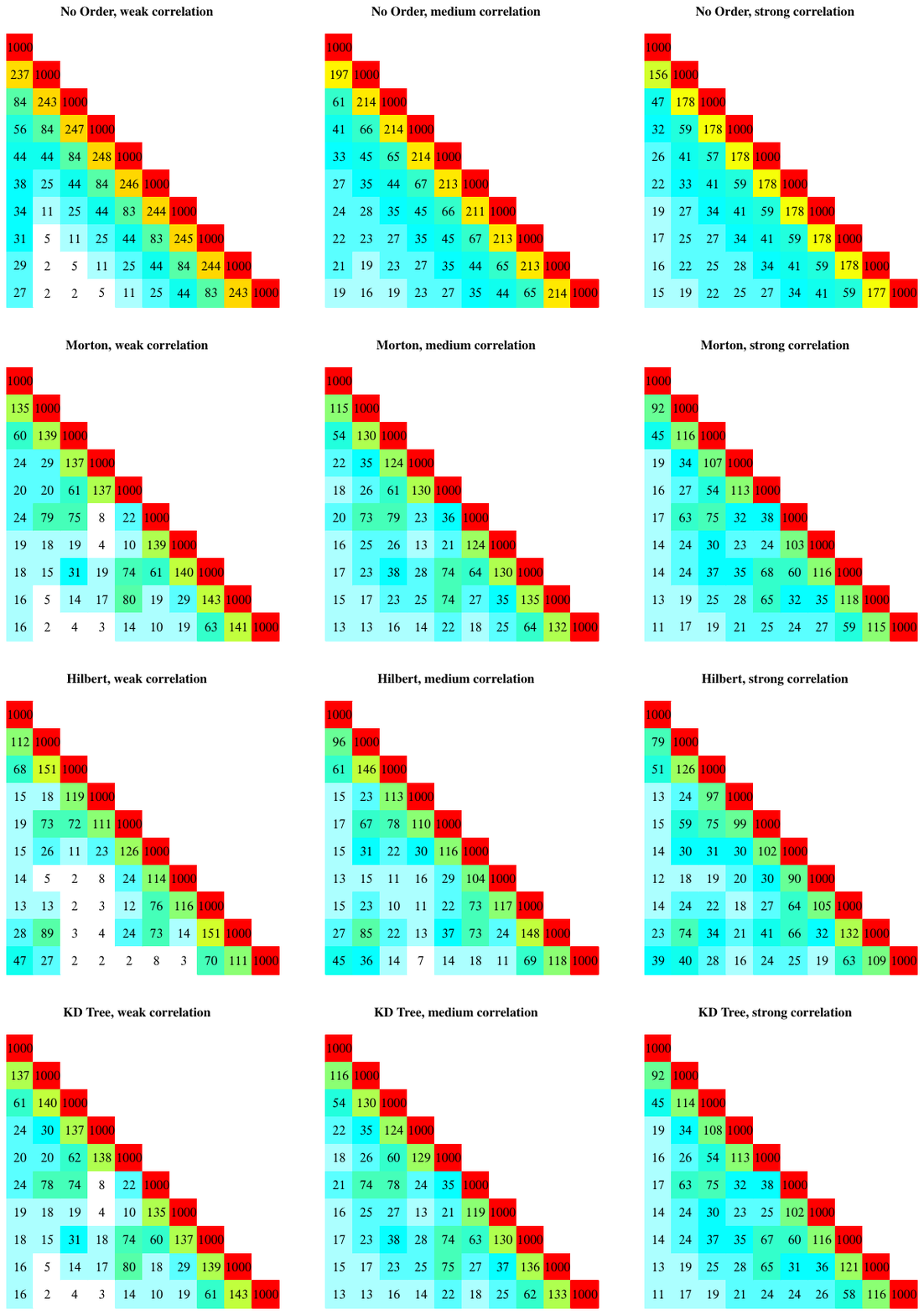}
    \caption{Heatmaps of tile ranks for various ordering algorithms for weak, medium, and strong correlation structures in smooth settings. Each small square symbolizes a $1,000 \times 1,000$ tile, annotated with its corresponding rank. Diagonal tiles maintain full rank. Within each heatmap, darker colours indicate higher ranks.}
    \label{rankplot_s}
\end{figure}

\begin{figure}[htbp]
    \centering
    \includegraphics[width=0.9\linewidth]{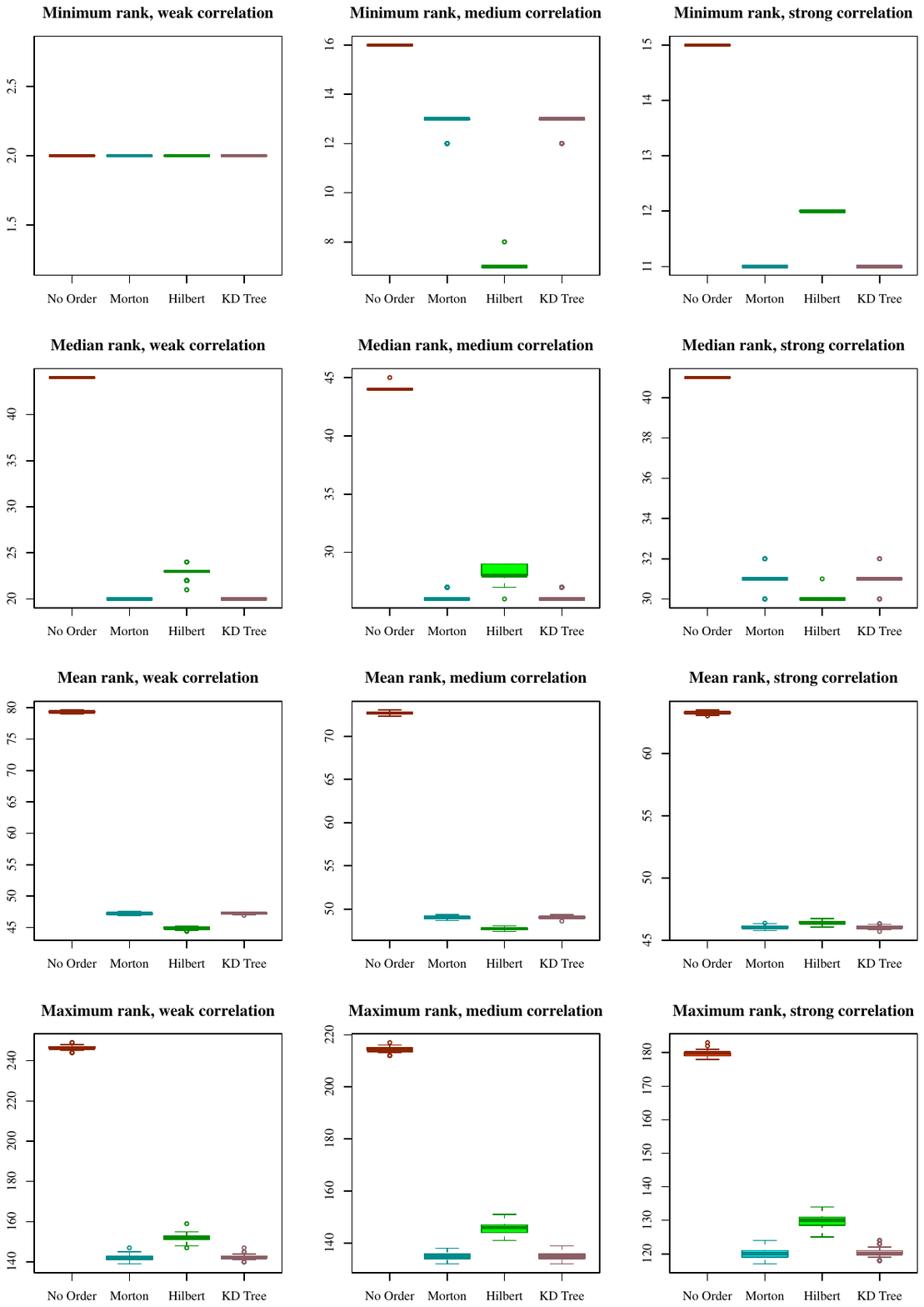}
    \caption{BoxPlots of the minimum, median, mean, and maximum of off-diagonal tile ranks with different ordering methods with weak, medium, and strong correlation structures in smooth settings. We generated $100$ sets of data with $n=10{,}000$ locations, and they are all divided into $1{,}000\times1{,}000$ tiles. }
    \label{rankboxplot_s}
\end{figure}

\begin{figure}[htbp]
    \centering
    \includegraphics[width=0.9\linewidth]{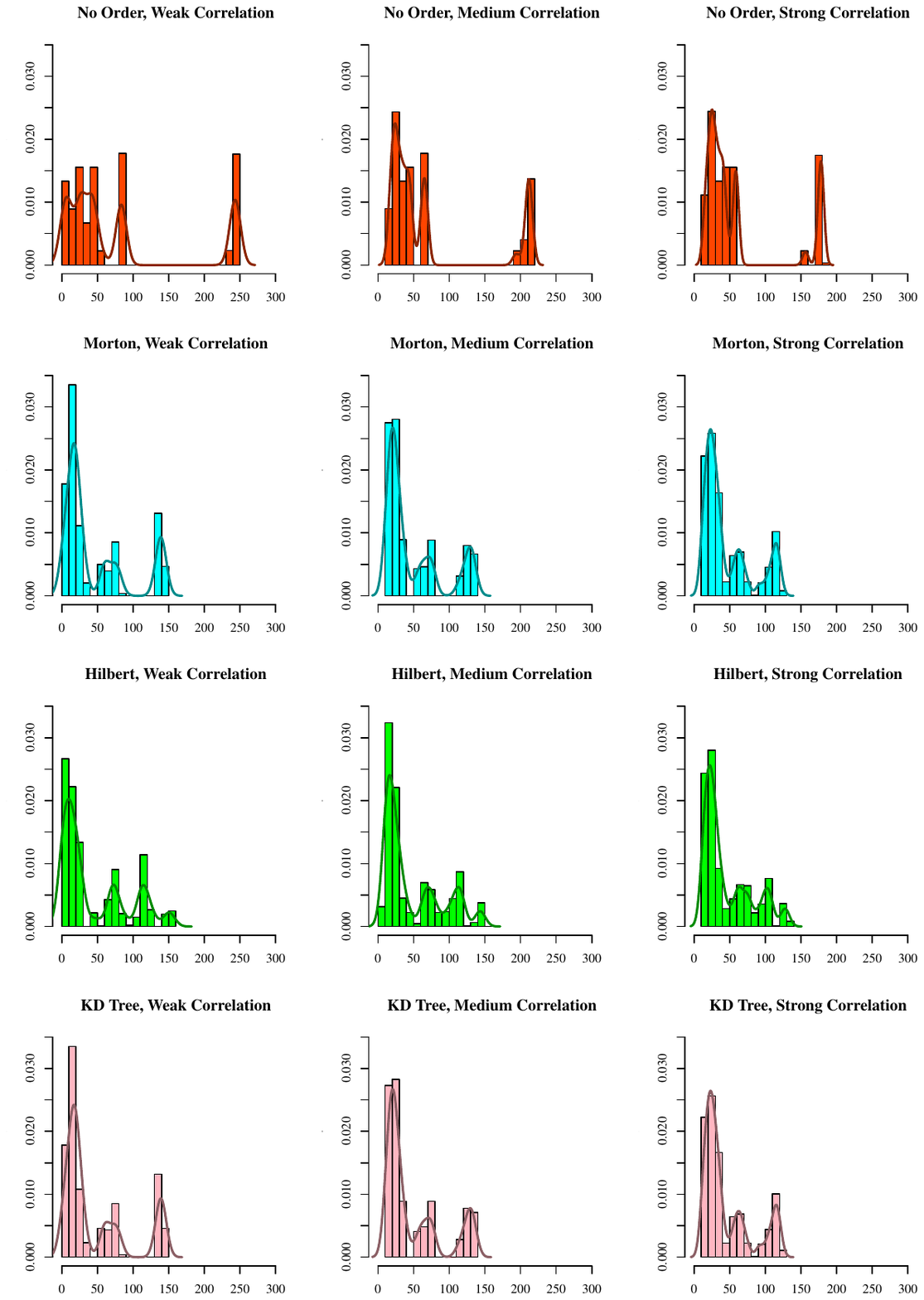}
    \caption{Histograms of the proportions and curves of the empirical densities of off-diagonal tile ranks with different ordering methods with weak, medium, and strong correlation structures in the smooth settings. We generated $100$ sets of data with $n=10{,}000$ locations, and they are all divided into $1{,}000\times1{,}000$ tiles.}
    \label{rankhist_s}
\end{figure}

As in Section \ref{ranks}, we show the heatmaps of some sample covariance matrices with weak, medium, and strong correlation structures in the smooth settings in Figure \ref{rankplot_s}. In addition, for each type of correlation structure, we generated 100 sets of spatial data and made boxplots for the minimum, median, mean, and maximum values of the off-diagonal tiles in the corresponding covariance matrices, which is shown in Figure \ref{rankboxplot_s}; we also show the histogram of these ranks from the 100 data sets, which we demonstrate in Figure \ref{rankhist_s}.

\end{document}